\documentclass[prc,twocolumn,superscriptaddress,showpacs,amssymb,amsmath,amsfonts,aps]{revtex4-1}
\usepackage{graphicx}
\usepackage{dcolumn}
\newcolumntype{d}[1]{D{.}{.}{-1}} 
\usepackage{epsfig}
\usepackage{latexsym}
\usepackage{amsmath} 
\usepackage{url}
\usepackage{natbib}
\usepackage[dvips]{color}

\begin{document}
\date{\today}
\title{\large Measurements of $e p \to e' \pi^+ \pi^- p'$ Cross Sections with CLAS at 1.40~GeV 
$<$ $W$ $<$ 2.0~GeV and 2.0~GeV$^2$ $< Q^2 <$ 5.0~GeV$^2$\\}

\newcommand*{\ANL}{Argonne National Laboratory, Argonne, Illinois 60439}
\newcommand*{\ANLindex}{1}
\affiliation{\ANL}
\newcommand*{\ASU}{Arizona State University, Tempe, Arizona 85287-1504}
\newcommand*{\ASUindex}{2}
\affiliation{\ASU}
\newcommand*{\CSUDH}{California State University, Dominguez Hills, Carson, CA 90747}
\newcommand*{\CSUDHindex}{3}
\affiliation{\CSUDH}
\newcommand*{\CANISIUS}{Canisius College, Buffalo, NY 14208}
\newcommand*{\CANISIUSindex}{4}
\affiliation{\CANISIUS}
\newcommand*{\CMU}{Carnegie Mellon University, Pittsburgh, Pennsylvania 15213}
\newcommand*{\CMUindex}{5}
\affiliation{\CMU}
\newcommand*{\CUA}{Catholic University of America, Washington, D.C. 20064}
\newcommand*{\CUAindex}{6}
\affiliation{\CUA}
\newcommand*{\SACLAY}{Irfu/SPhN, CEA, Universit\'e Paris-Saclay, 91191 Gif-sur-Yvette, France}
\newcommand*{\SACLAYindex}{7}
\affiliation{\SACLAY}
\newcommand*{\CNU}{Christopher Newport University, Newport News, Virginia 23606}
\newcommand*{\CNUindex}{8}
\affiliation{\CNU}
\newcommand*{\UCONN}{University of Connecticut, Storrs, Connecticut 06269}
\newcommand*{\UCONNindex}{9}
\affiliation{\UCONN}
\newcommand*{\FU}{Fairfield University, Fairfield CT 06824}
\newcommand*{\FUindex}{10}
\affiliation{\FU}
\newcommand*{\FERRARAU}{Universita' di Ferrara, 44121 Ferrara, Italy}
\newcommand*{\FERRARAUindex}{11}
\affiliation{\FERRARAU}
\newcommand*{\FIU}{Florida International University, Miami, Florida 33199}
\newcommand*{\FIUindex}{12}
\affiliation{\FIU}
\newcommand*{\FSU}{Florida State University, Tallahassee, Florida 32306}
\newcommand*{\FSUindex}{13}
\affiliation{\FSU}
\newcommand*{\Genova}{Universit$\grave{a}$ di Genova, 16146 Genova, Italy}
\newcommand*{\Genovaindex}{14}
\affiliation{\Genova}
\newcommand*{\GWUI}{The George Washington University, Washington, DC 20052}
\newcommand*{\GWUIindex}{15}
\affiliation{\GWUI}
\newcommand*{\ISU}{Idaho State University, Pocatello, Idaho 83209}
\newcommand*{\ISUindex}{16}
\affiliation{\ISU}
\newcommand*{\INFNFE}{INFN, Sezione di Ferrara, 44100 Ferrara, Italy}
\newcommand*{\INFNFEindex}{17}
\affiliation{\INFNFE}
\newcommand*{\INFNFR}{INFN, Laboratori Nazionali di Frascati, 00044 Frascati, Italy}
\newcommand*{\INFNFRindex}{18}
\affiliation{\INFNFR}
\newcommand*{\INFNGE}{INFN, Sezione di Genova, 16146 Genova, Italy}
\newcommand*{\INFNGEindex}{19}
\affiliation{\INFNGE}
\newcommand*{\INFNRO}{INFN, Sezione di Roma Tor Vergata, 00133 Rome, Italy}
\newcommand*{\INFNROindex}{20}
\affiliation{\INFNRO}
\newcommand*{\INFNTUR}{INFN, Sezione di Torino, 10125 Torino, Italy}
\newcommand*{\INFNTURindex}{21}
\affiliation{\INFNTUR}
\newcommand*{\ORSAY}{Institut de Physique Nucl\'eaire, CNRS/IN2P3 and Universit\'e Paris Sud, Orsay, France}
\newcommand*{\ORSAYindex}{22}
\affiliation{\ORSAY}
\newcommand*{\Juelich}{Institute fur Kernphysik (Juelich), 48149 Juelich, Germany}
\newcommand*{\Juelichindex}{23}
\affiliation{\Juelich}
\newcommand*{\ITEP}{Institute of Theoretical and Experimental Physics, Moscow, 117259, Russia}
\newcommand*{\ITEPindex}{24}
\affiliation{\ITEP}
\newcommand*{\JMU}{James Madison University, Harrisonburg, Virginia 22807}
\newcommand*{\JMUindex}{25}
\affiliation{\JMU}
\newcommand*{\KNU}{Kyungpook National University, Daegu 41566, Republic of Korea}
\newcommand*{\KNUindex}{26}
\affiliation{\KNU}
\newcommand*{\MISS}{Mississippi State University, Mississippi State, MS 39762-5167}
\newcommand*{\MISSindex}{27}
\affiliation{\MISS}
\newcommand*{\UNH}{University of New Hampshire, Durham, New Hampshire 03824-3568}
\newcommand*{\UNHindex}{28}
\affiliation{\UNH}
\newcommand*{\NSU}{Norfolk State University, Norfolk, Virginia 23504}
\newcommand*{\NSUindex}{29}
\affiliation{\NSU}
\newcommand*{\OHIOU}{Ohio University, Athens, Ohio  45701}
\newcommand*{\OHIOUindex}{30}
\affiliation{\OHIOU}
\newcommand*{\ODU}{Old Dominion University, Norfolk, Virginia 23529}
\newcommand*{\ODUindex}{31}
\affiliation{\ODU}
\newcommand*{\RPI}{Rensselaer Polytechnic Institute, Troy, New York 12180-3590}
\newcommand*{\RPIindex}{32}
\affiliation{\RPI}
\newcommand*{\URICH}{University of Richmond, Richmond, Virginia 23173}
\newcommand*{\URICHindex}{33}
\affiliation{\URICH}
\newcommand*{\ROMAII}{Universita' di Roma Tor Vergata, 00133 Rome Italy}
\newcommand*{\ROMAIIindex}{34}
\affiliation{\ROMAII}
\newcommand*{\MSU}{Skobeltsyn Institute of Nuclear Physics and Physics Department, Lomonosov Moscow State University, 119234 Moscow, Russia}
\newcommand*{\MSUindex}{35}
\affiliation{\MSU}
\newcommand*{\SCAROLINA}{University of South Carolina, Columbia, South Carolina 29208}
\newcommand*{\SCAROLINAindex}{36}
\affiliation{\SCAROLINA}
\newcommand*{\TEMPLE}{Temple University,  Philadelphia, PA 19122}
\newcommand*{\TEMPLEindex}{37}
\affiliation{\TEMPLE}
\newcommand*{\JLAB}{Thomas Jefferson National Accelerator Facility, Newport News, Virginia 23606}
\newcommand*{\JLABindex}{38}
\affiliation{\JLAB}
\newcommand*{\UTFSM}{Universidad T\'{e}cnica Federico Santa Mar\'{i}a, Casilla 110-V Valpara\'{i}so, Chile}
\newcommand*{\UTFSMindex}{39}
\affiliation{\UTFSM}
\newcommand*{\EDINBURGH}{Edinburgh University, Edinburgh EH9 3JZ, United Kingdom}
\newcommand*{\EDINBURGHindex}{40}
\affiliation{\EDINBURGH}
\newcommand*{\GLASGOW}{University of Glasgow, Glasgow G12 8QQ, United Kingdom}
\newcommand*{\GLASGOWindex}{41}
\affiliation{\GLASGOW}
\newcommand*{\VT}{Virginia Tech, Blacksburg, Virginia 24061-0435}
\newcommand*{\VTindex}{42}
\affiliation{\VT}
\newcommand*{\VIRGINIA}{University of Virginia, Charlottesville, Virginia 22901}
\newcommand*{\VIRGINIAindex}{43}
\affiliation{\VIRGINIA}
\newcommand*{\WM}{College of William and Mary, Williamsburg, Virginia 23187-8795}
\newcommand*{\WMindex}{44}
\affiliation{\WM}
\newcommand*{\YEREVAN}{Yerevan Physics Institute, 375036 Yerevan, Armenia}
\newcommand*{\YEREVANindex}{45}
\affiliation{\YEREVAN}

\newcommand*{\NOWJLAB}{Thomas Jefferson National Accelerator Facility, Newport News, Virginia 23606}
\newcommand*{\NOWHAMPTON}{Hampton University, Hampton, VA 23668}
\newcommand*{\NOWISU}{Idaho State University, Pocatello, Idaho 83209}
\newcommand*{\NOWGLASGOW}{University of Glasgow, Glasgow G12 8QQ, United Kingdom}
\newcommand*{\NOWINFNGE}{INFN, Sezione di Genova, 16146 Genova, Italy}

\author {E.~L.~Isupov} 
\affiliation{\MSU}
\author {V.~D.~Burkert} 
\affiliation{\JLAB}
\author {D.~S.~Carman} 
\affiliation{\JLAB}
\author {R.~W.~Gothe}
\affiliation{\SCAROLINA}
\author {K.~Hicks} 
\affiliation{\OHIOU}
\author {B.~S.~Ishkhanov} 
\affiliation{\MSU}
\author {V.~I.~Mokeev} 
\affiliation{\JLAB}
\author {K.P. ~Adhikari} 
\affiliation{\MISS}
\author {S. Adhikari} 
\affiliation{\FIU}
\author {D.~Adikaram} 
\altaffiliation[Current address: ]{\NOWJLAB}
\affiliation{\ODU}
% ...

\collaboration{The CLAS Collaboration}
\noaffiliation

\begin{abstract}
{This paper reports new exclusive cross sections for $ep \to e' \pi^+\pi^-p'$ using 
the CLAS detector at Jefferson Laboratory. These results are presented for the 
first time at photon virtualities 2.0~GeV$^2$ $< Q^2 < 5.0$~GeV$^2$ in the center-of-mass 
energy range 1.4~GeV $< W <$ 2.0~GeV, which covers a large part of the nucleon 
resonance region. Using a model developed for the phenomenological analysis of
electroproduction data, we see strong indications that the relative contributions 
from the resonant cross sections at $W < 1.74$~GeV increase with $Q^2$. These data 
considerably extend the kinematic reach of previous measurements. Exclusive 
$ep \to e' \pi^+\pi^-p'$ cross section measurements are of particular importance for 
the extraction of resonance electrocouplings in the mass range above 1.6~GeV.}
\end{abstract}


\maketitle

\section{Introduction}
\label{intro}

An extensive research program aimed at the exploration of the structure of excited 
nucleon states is in progress at Jefferson Lab, employing exclusive meson 
electroproduction off protons in the nucleon resonance ($N^*$) region. This
represents an important direction in a broad effort to analyze data from the CLAS 
detector~\cite{Bu12,Az13,Bu15}. 

Many nucleon states in the mass range above 1.6~GeV are known to couple strongly to
$\pi\pi N$. Therefore, studies of exclusive $\pi^+\pi^-p$ electroproduction are a 
major source of information on the internal structure of these states. Studies of 
exclusive $\pi^+\pi^-p$ electroproduction are of particular importance for the 
extraction of the $N^*$ electrocoupling amplitudes off protons for all prominent 
resonances in the mass range up to 2.0~GeV and at photon virtualities 
$Q^2 < 5.0$~GeV$^2$. 

The $\gamma_vpN^*$ electrocouplings are the primary source of information on many 
facets of non-perturbative strong interactions, particularly in the generation of 
the excited proton states from quarks and gluons. Analyses of the $\gamma_vpN^*$ 
electrocouplings extracted from CLAS have already revealed distinctive differences 
in the electrocouplings of states with different underlying quark structures, 
{\it e.g.} orbital versus radial quark excitations~\cite{Bu12,Az13,Bu15}.   

Furthermore, the structure of excited nucleons represents a complex interplay between 
the inner core of three dressed quarks and the external meson-baryon cloud 
\cite{Bu12,Mo16,Az15,Lee10}, with their relative contributions evolving with photon 
virtuality. Therefore, measurements of $\gamma_vpN^*$ electrocouplings allow for 
a detailed charting of the spatial structure of nucleon resonances in terms of their 
quark cores and higher Fock states. Studies of many prominent resonances are needed 
in order to explore the full complexity of non-perturbative strong interactions in 
the generation of different excited states. It is through such information that models 
built on ingredients from QCD are to be confronted, and lead to new insights into the 
strong interaction dynamics, as well as developments of new theoretical approaches to 
solve QCD in these cases. 

The unique interaction of experiment and theory was recently demonstrated on the quark 
distribution amplitudes (DAs) for the $N(1535)1/2^-$ resonance (a chiral partner of 
the nucleon ground state). These DAs have become available from Lattice QCD~\cite{Br14}, 
constrained by the CLAS results on the transition $N \to N(1535)1/2^-$ form factor 
\cite{Az09}, by employing DAs and the Light Cone Sum Rule (LCSR) approach~\cite{Br15}. 
The comparison of quark DAs in the nucleon ground state and in the $N(1535)1/2^-$ 
resonance demonstrates a pronounced difference, elucidating the manifestation of 
Dynamical Chiral Symmetry Breaking (DCSB) in the structure of the ground and excited 
nucleon states. 

Recent advances in Dyson-Schwinger Equations (DSEs) now make it possible to describe 
the elastic nucleon and the transition form factors for $N \to \Delta(1232)3/2^+$ 
and $N \to N(1440)1/2^+$ starting from the QCD Lagrangian~\cite{Seg14,Seg15}. 
Currently, DSEs relate the $\gamma_vpN^*$ electrocouplings to the quark mass function 
at distance scales of $Q^2 > 2$~GeV$^2$, where the quark core is the biggest contributor 
to the $N^*$ structure. This success demonstrates the relevance of dressed constituent 
quarks inferred within the DSEs \cite{Cr14} as effective degrees of freedom in the 
structure of the ground and excited nucleon states, and emphasizes the need for data on 
the $Q^2$ dependence of the $\gamma_vpN^*$ electrocouplings to provide access to the 
momentum dependence of the dressed quark mass. This can provide new insight into 
two of the still open problems of the Standard Model, namely the nature of hadron mass 
and the emergence of quark-gluon confinement from QCD~\cite{Cr16,Cr16a,Cr14}.  

The CLAS Collaboration has provided much of the world data on meson electroproduction 
in the resonance excitation region. Nucleon resonance electrocouplings have been obtained 
from the exclusive channels: $\pi^+n$ and $\pi^0p$ at $Q^2 < 5.0$~GeV$^2$ in the mass 
range up to 1.7~GeV, $\eta p$ at $Q^2 < 4.0$~GeV$^2$ in the mass range up to 1.6~GeV, 
and $\pi^+\pi^-p$ at $Q^2 < 1.5$~GeV$^2$ in the mass range up to 1.8~GeV 
\cite{Mo16,Bu12,Az09,Park15,Mo12,Mo14,Mo16a,Mo16b}. The studies of the $N(1440)1/2^+$ 
and $N(1520)3/2^-$ resonances with the CLAS detector~\cite{Mo16,Az09,Mo12} have provided 
most of the information available worldwide on these electrocouplings in the range  
0.25~GeV$^2 < Q^2 < 5.0$~GeV$^2$. The $N(1440)1/2^+$ and $N(1520)3/2^-$ states, together 
with the $\Delta(1232)3/2^+$ and $N(1535)1/2^-$ resonances, are the best understood 
excited nucleon states to date~\cite{Bu12}. Furthermore, results on the $\gamma_{v}pN^*$ 
electrocouplings for the high-lying $N(1675)5/2^-$, $N(1680)5/2^+$, and $N(1710)1/2^+$ 
resonances were determined from the CLAS $\pi^+ n$ data at 1.5~GeV$^2 < Q^2 < 4.5$~GeV$^2$
\cite{Park15}. 

Many excited nucleon states with masses above 1.6~GeV decay preferentially to the 
$\pi\pi N$ final states, making exclusive $\pi^+\pi^-p$ electroproduction off protons a 
major source of information on these electrocouplings. First accurate results on the 
electrocouplings of the $\Delta(1620)1/2^-$, which couples strongly to $\pi\pi N$, 
have been published from the analysis of CLAS data on $\pi^+\pi^-p$ electroproduction 
off protons~\cite{Mo16}. Preliminary results on electrocouplings of two other 
resonances, the $\Delta(1700)3/2^-$ and the $N(1720)3/2^+$, show dominance of 
$\pi\pi N$ decays and were obtained from the $\pi^+\pi^-p$ data~\cite{Mo14}. Previous 
studies of these resonances in the $\pi N$ final states suffered from large uncertainties 
due to small branching fractions for decays to $\pi N$.

The combined analysis of the $\pi^+\pi^-p$ photo- and electroproduction data~\cite{Ri03} 
revealed preliminary evidence for the existence of a $N'(1720)3/2^+$ state. Its 
spin-parity, mass, total and partial hadronic decay widths, along with the $Q^2$
evolution of its $\gamma_vpN^*$ electrocouplings, have been obtained from a fit to 
the CLAS data~\cite{Mo16a}. This is the only new candidate state for which information 
on $\gamma_vpN^*$ electrocouplings has become available, offering access to its 
internal structure. A successful description of the photo- and electroproduction data 
with $Q^2$ independent mass and hadronic decay widths offers nearly model-independent 
evidence for the existence of this state. Future studies of exclusive $\pi^+\pi^-p$ 
electroproduction off protons at $W > 1.7$~GeV will also open up the possibility to 
verify new baryon states observed in a global multi-channel analysis of exclusive 
photoproduction data by the Bonn-Gatchina group~\cite{Sar12}.  

The resonance electrocouplings from exclusive $\pi^+\pi^-p$ electroproduction off 
protons have been extracted in the range of $W < 2.0$~GeV and $Q^2 < 1.5$~GeV$^2$ 
\cite{Ri03,Fe09}. An extension of the measured $\pi^+\pi^-p$ electroproduction cross 
sections towards higher photon virtualities is critical for the extraction of 
resonance electrocouplings at the distance scale where the transition to the 
dominance of dressed quark degrees of freedom in the $N^*$ structure is expected 
\cite{Bu12,Az13}. These data will provide input for reaction models aimed at 
determining $\gamma_vpN^*$ electrocouplings for the $N^*$ resonances in the mass 
range above 1.6~GeV~\cite{Mo09,Mo12,Mo16}. These data will also provide necessary 
input for global multi-channel analyses of the exclusive meson photo-, electro-, and 
hadroproduction channels~\cite{Sar12,Sar14,Lee10,Lee13,Lee15}. 

In this paper we present cross sections for $\pi^+\pi^-p$ electroproduction off 
protons at center of mass energies $W$ from 1.4~GeV to 2.0~GeV and at $Q^2$ from 
2.0~GeV$^2$ to 5.0~GeV$^2$ in terms of nine independent 1-fold differential and 
fully integrated $\pi^+\pi^-p$ cross sections. As in our previous studies
\cite{Ri03,Fe09}, these are obtained by integration of the 5-fold differential 
cross section over different sets of four kinematic variables. The combined 
analysis of all nine 1-fold differential cross sections gives access to correlations 
in the 5-fold differential cross sections from the correlations seen in the nine 
1-fold differential cross sections, as they all represent different integrals of 
the same 5-fold differential cross sections. 
 
\section{Experimental Description}
\label{expt}

The data were collected using the CLAS detector \cite{Me03} with an electron beam of 
5.754~GeV incident on a liquid-hydrogen target. The beam current averaged about 7~nA 
and was produced by the Continuous Electron Beam Accelerator Facility (CEBAF) at the 
Thomas Jefferson National Accelerator Laboratory (TJNAF). The liquid-hydrogen target 
had a length of 5.0~cm and was placed 4.0~cm upstream of the center of the CLAS 
detector. The torus coils of the CLAS detector were operated at 3375~A and an additional 
mini-torus close to the target was run at 6000~A to remove low-energy background 
electrons. The CLAS spectrometer consisted of a series of detectors in each of its
six azimuthal sectors, including three sets of wire drift-chambers (DC) for tracking 
scattered charged particles, Cerenkov counters (CC) to distinguish electrons and 
pions, sampling electromagnetic calorimeters (EC) for electron and neutral particle
identification, and a set of time-of-flight scintillation counters (SC) to record the 
flight time of charged particles. For this experiment, the data acquisition triggered 
on a coincidence between signals in the CC and EC, as explained below. This configuration 
of the experiment was called the CLAS e1-6 run to distinguish it from other data sets.

\subsection{Selection of Electrons}
\label{sel}

\begin{figure*}[htp]
\begin{center}
\includegraphics[width=13cm,keepaspectratio]{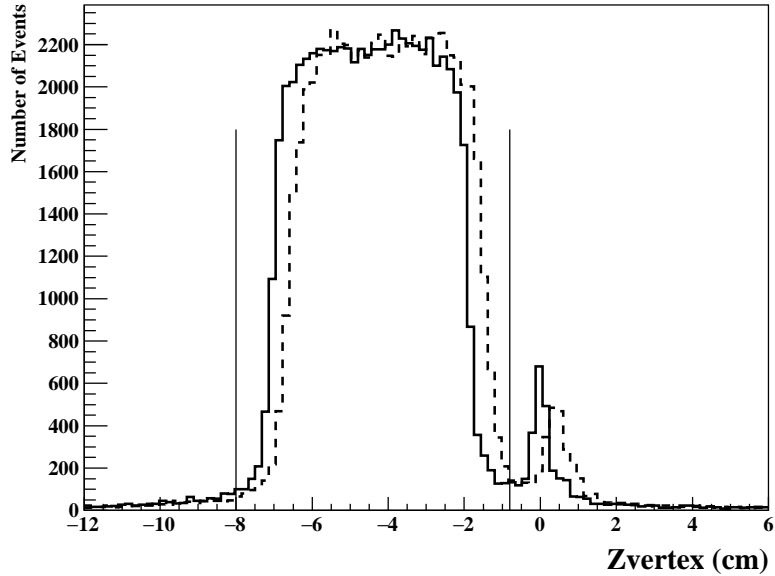}
\vspace{-0.1cm}
\caption{ Vertex reconstruction projected onto the beam axis for Sector~2 of CLAS, 
before (dashed) and after (black) applying corrections to align the sectors of CLAS. 
The vertical lines show the region of the vertex event selection. The small peak at 
zero originates from an aluminum window 2 cm downstream of the target cell.}
\label{fig:zvertex}
\end{center}
\end{figure*}

The particle tracks were determined from the DC coordinates and extrapolated back to 
the target position. A coordinate system was defined with the $z$-axis along the beam 
direction. A histogram of a sampling of electron tracks extrapolated to their point of 
closest approach to the $z$-axis is shown in Fig.~\ref{fig:zvertex} for one of the six 
sectors of the CLAS detector. Plots for the other sectors are very similar. A small 
correction was made for the positioning of the DC in each sector to align the target 
position. Event selection required a good event to come from the target region.

\begin{figure*}[htp]
\begin{center}
\includegraphics[width=10cm,keepaspectratio]{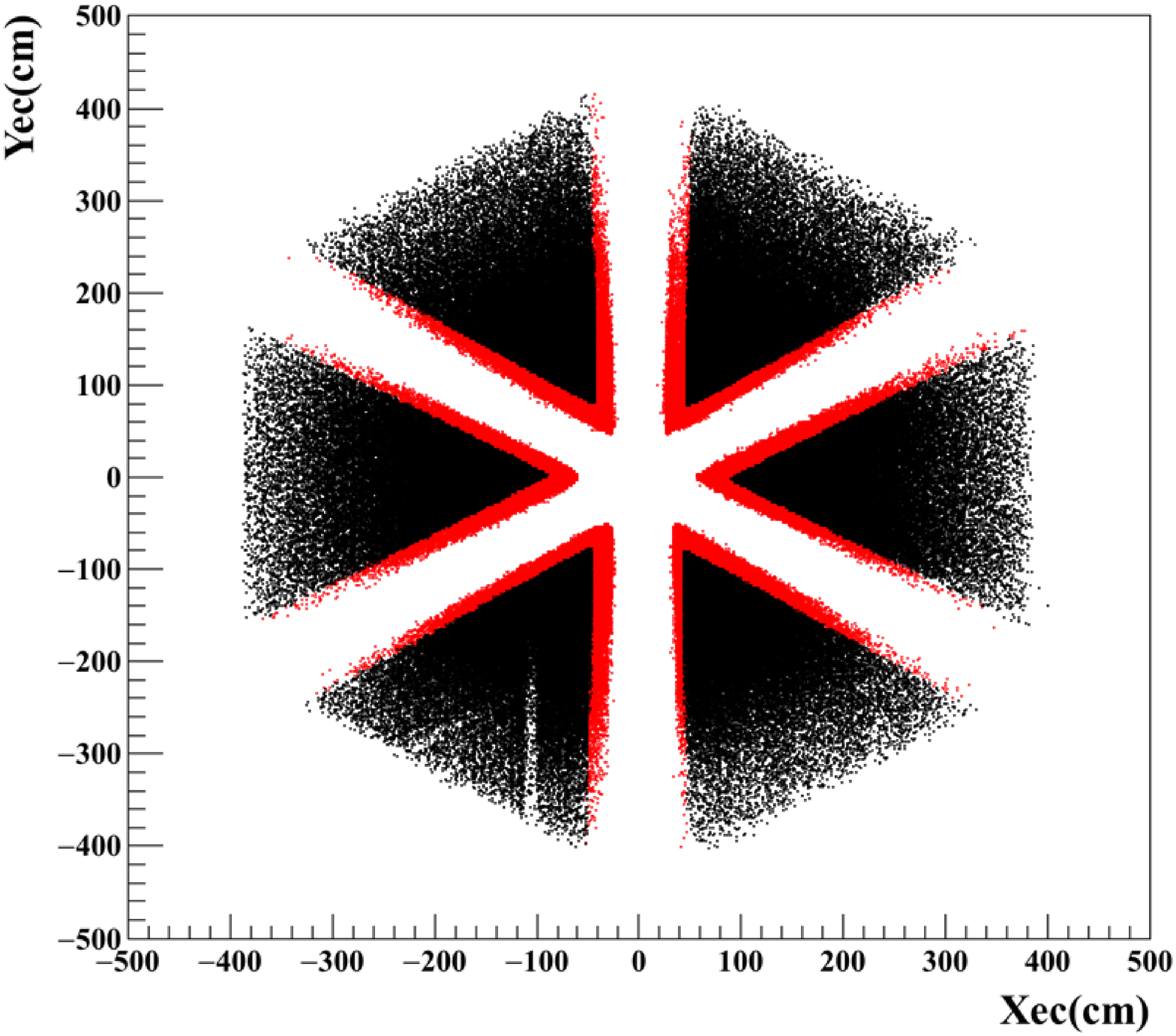}
\vspace{-0.1cm}
\caption{(Color Online) The position of electron events in the EC for the six sectors 
of CLAS for all events (light gray or red online) and selected events (black). The 
stripe seen in the lower left sector is due to inefficient phototubes on a few 
scintillator strips of the EC. The same inefficiencies are introduced in the simulations 
of the detector acceptance.}
\label{fig:fiducial}
\end{center}
\end{figure*}

A scattered electron produced an electromagnetic shower of particles in the EC, and 
the characteristics of this shower were different for pions and electrons. However, 
the electromagnetic shower was not fully contained at the edges of the EC, 
so it was necessary to place an event selection cut to remove these unwanted events 
near the edges.  This cut on the fiducial volume is shown in Fig.~\ref{fig:fiducial}. 
The edges of the fiducial regions were chosen based on studies of the EC resolution 
and the comparison with known cross sections for elastic $e-p$ scattering.

\begin{figure*}[htp]
\begin{center}
\includegraphics[width=13cm,keepaspectratio]{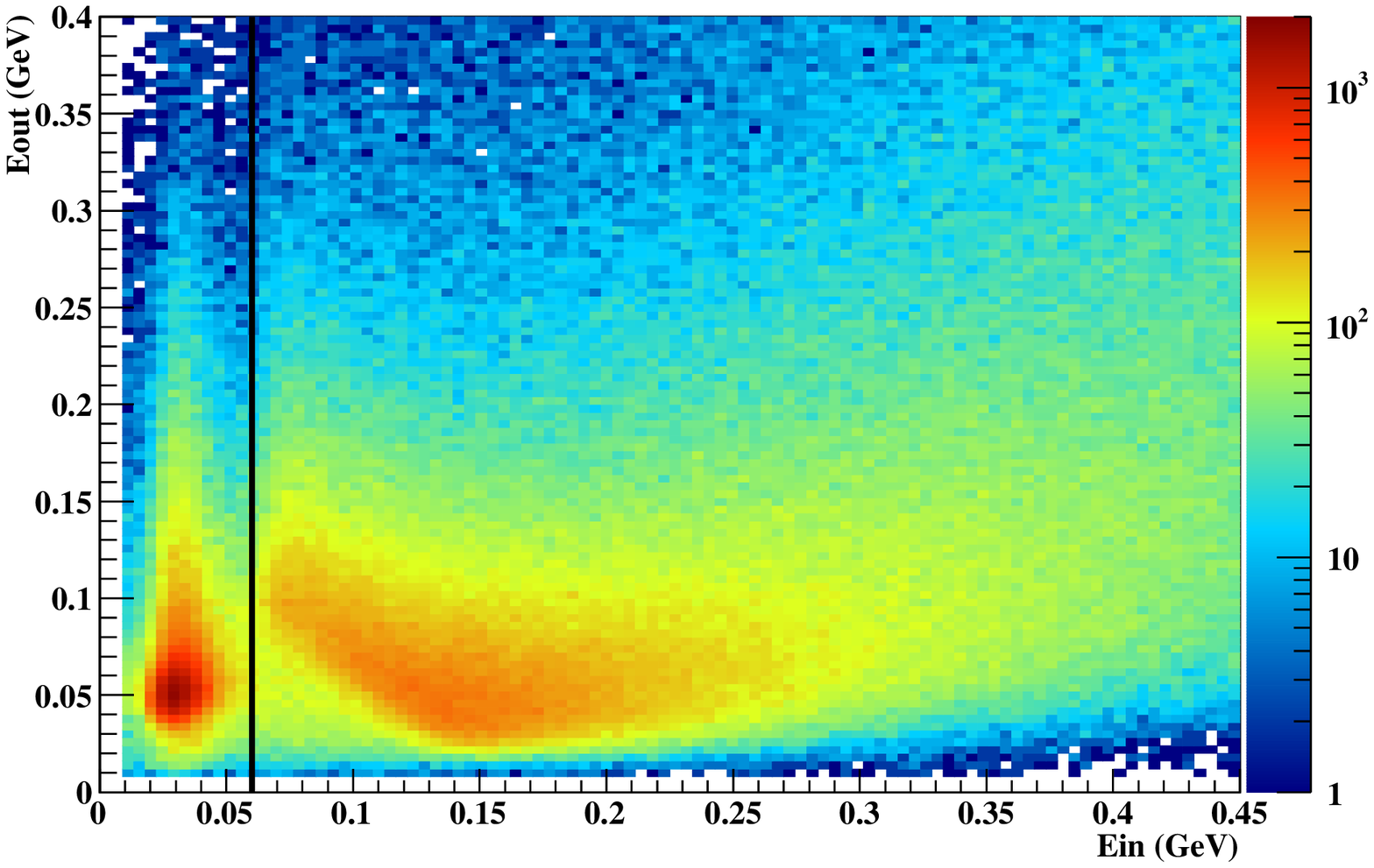}
\vspace{-0.1cm}
\caption{(Color Online) The energy deposited in the inner ($E_{in}$) and outer ($E_{out}$) 
layers of the EC for all particles. The line corresponds to 60~MeV, which separates the 
minimum ionizing pions (to the left) and electrons (to the right).}
\label{fig:ecinout}
\end{center}
\end{figure*}

The EC has two layers, an inner layer (closer to the target) and an outer layer. See 
Ref.~\cite{Me03} for more details on the EC geometry. The two layers enabled separation 
of charged pions and electrons. Normally incident minimum ionizing pions typically lost 
26~MeV of energy in the 15~cm of scintillating material of the inner part of the 
calorimeter, whereas electrons that underwent an electromagnetic shower, deposited more 
energy ($E_{in}$) in the inner EC layer. A data selection cut $E_{in} > 60$~MeV eliminated 
most of the pions, as shown in Fig.~\ref{fig:ecinout}. A further refined selection of 
electrons came from the correlation between total energy deposited and momentum. An 
additional momentum-dependent cut was placed on the ratio of the total energy in the EC 
and the momentum, $E_{tot}/p$. For a given momentum, the data formed a Gaussian peak for 
this ratio centered near 0.3. A 2.5$\sigma$ cut on this peak was applied to the data.  
The loss of events in the Gaussian tail was accounted for by the detector acceptance, 
where an equivalent cut was placed on the Monte Carlo simulation data.

\subsection{Particle Identification}
\label{iden}

\begin{figure*}[htp]
\begin{center}
\includegraphics[width=10cm,keepaspectratio]{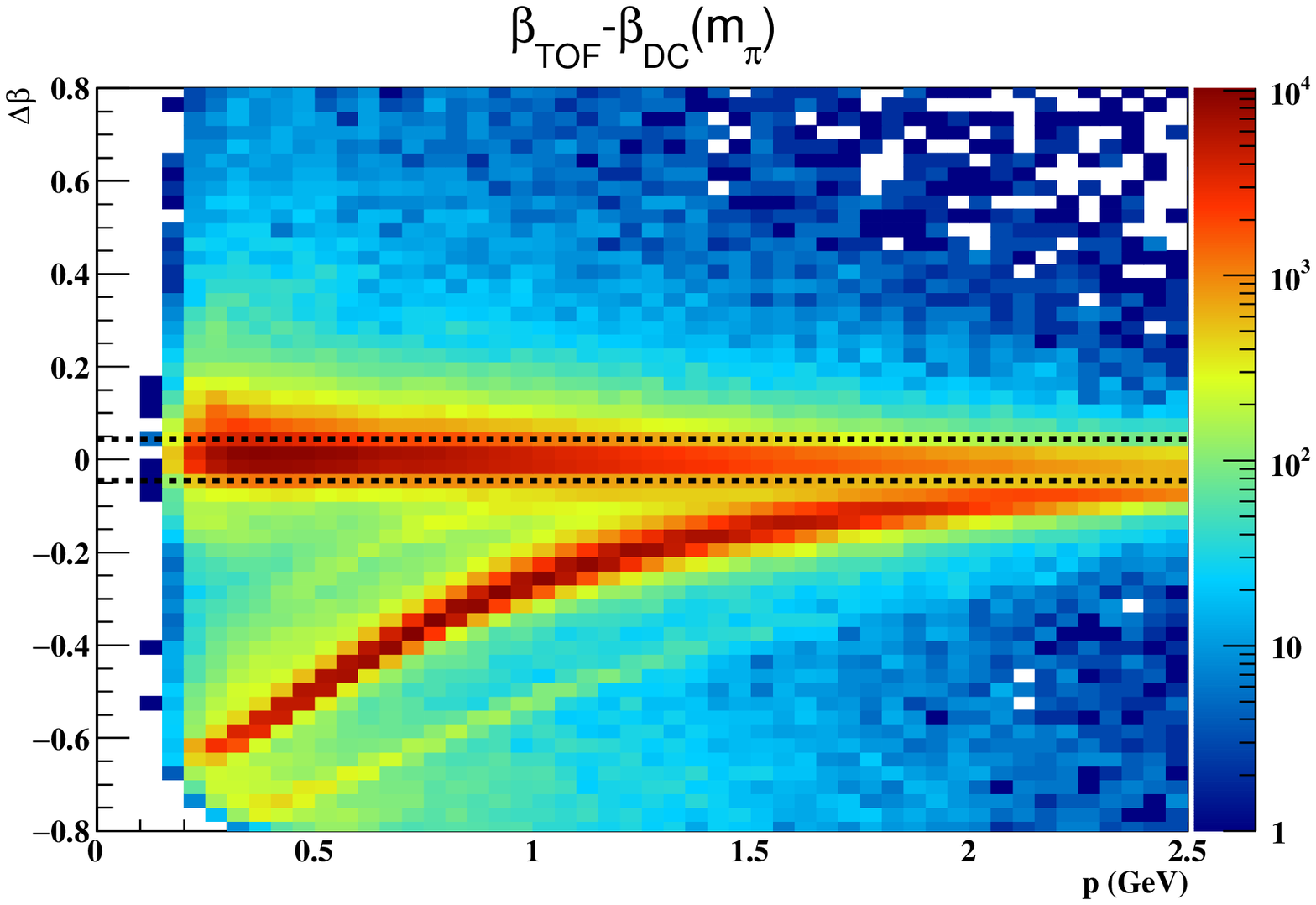}
\includegraphics[width=10cm,keepaspectratio]{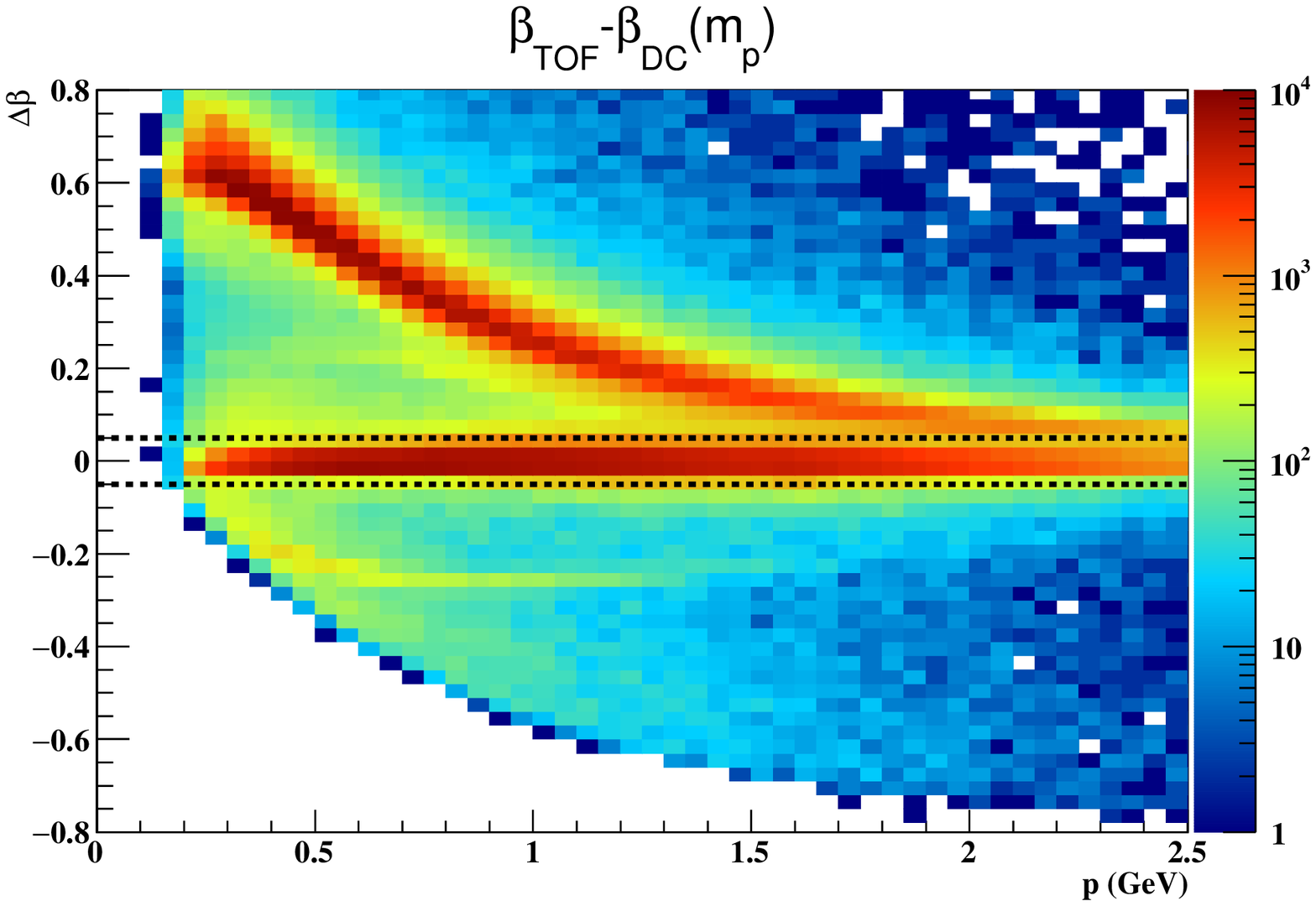}
\vspace{-0.1cm}
\caption{(Color Online) Velocity difference $\beta_{TOF}-\beta_{DC}$ for a sample of
positively charged tracks versus momentum for assumed masses of a pion (top) or a proton 
(bottom).}
\label{fig:deltabeta}
\end{center}
\end{figure*}

Particle identification for hadrons was obtained comparing the particle velocity 
evaluated from the flight time (from the target to the SC) and from the momentum of 
the particle track (measured by the DC) for an assumed mass. When the assumed 
particle mass is correct, the particle's velocity calculated from both methods agrees. 
Fig.~\ref{fig:deltabeta} top and bottom show the difference between the velocity 
calculated from the time-of-flight and that from the momentum for pions and protons, 
respectively, which gives a horizontal band about zero velocity difference. Below a 
momentum of about 2~GeV, this method provides excellent separation between pions and 
protons, and reasonable separation up to 2.5~GeV. 

\begin{figure*}[htp]
\begin{center}
\includegraphics[width=10cm,keepaspectratio]{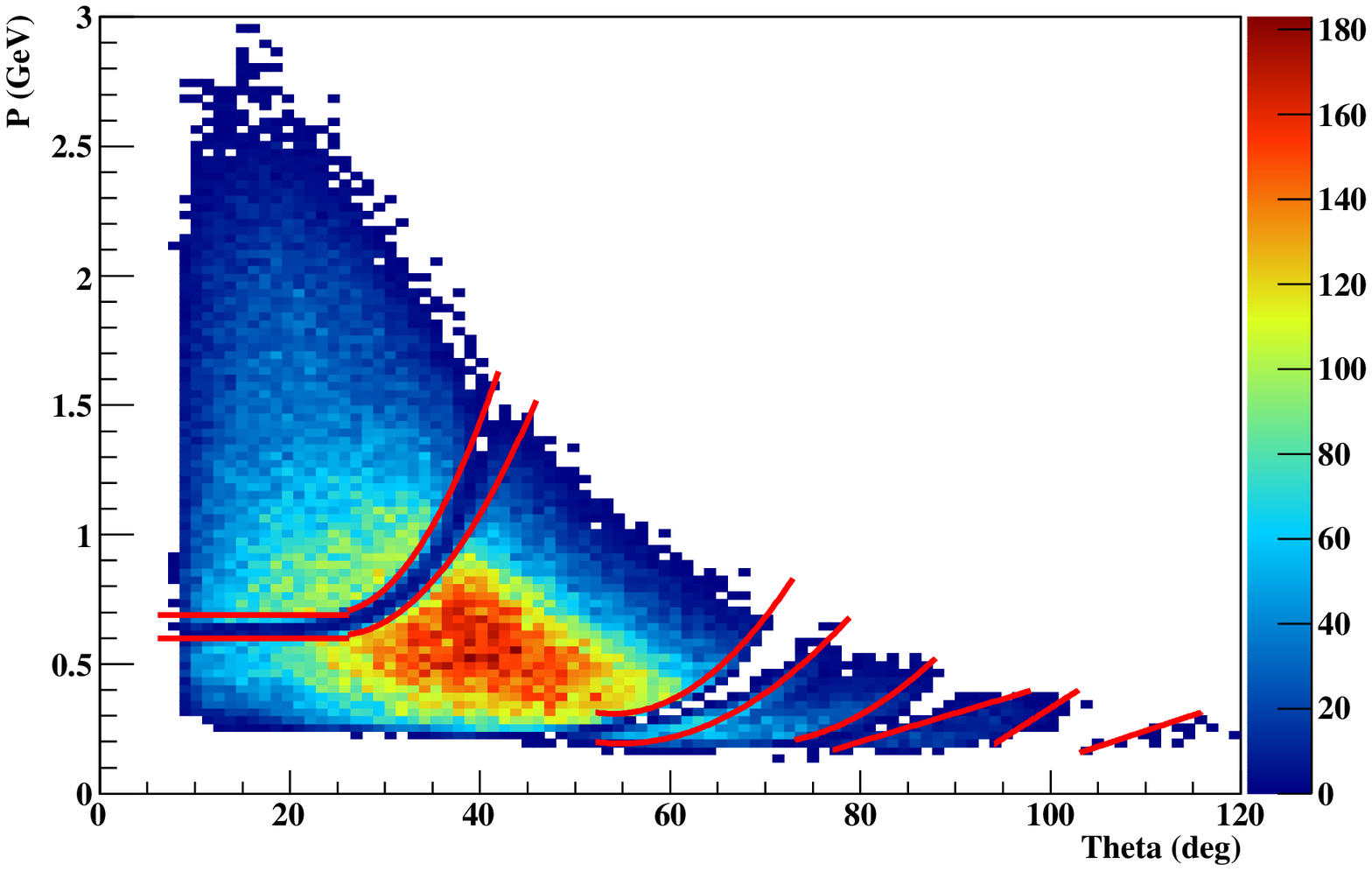}
\vspace{-0.1cm}
\caption{(Color Online) Histogram of the correlation between the momentum $p$ and the
polar angle $\theta$ for tracks of positively charged pions in Sector~3 of CLAS. The
inefficient regions of the detector, shown between the bands of solid lines, were removed
from the analysis.}
\label{fig:sector3}
\end{center}
\end{figure*}

For the e1-6 run, the current in the torus coils was set such that positively charged 
particles bent outward and negatively charged particles bent inward. In this data run, 
some regions of the CLAS detector were inefficient, due to bad sections of the DC or bad SC 
paddle PMTs. An example is shown in Fig.~\ref{fig:sector3} for positively charged pions in 
Sector~3. The inefficient detector regions show up clearly in a plot of the measured track 
momentum $p$ versus the polar angle $\theta$ of the track. These regions were cut out of both 
the data and Monte Carlo simulation, providing a good match between the real and simulated 
detector acceptance. In addition, cuts were placed to restrict particle tracks to the 
fiducial volume of the detector, which eliminated inefficient regions at the edges of the 
CC and DC. The fiducial cuts are standard for CLAS and are described elsewhere~\cite{Ri03}. 

\subsection{Event Selection}
\label{evsel}

\begin{figure*}[htp]
\begin{center}
\includegraphics[width=10cm,keepaspectratio]{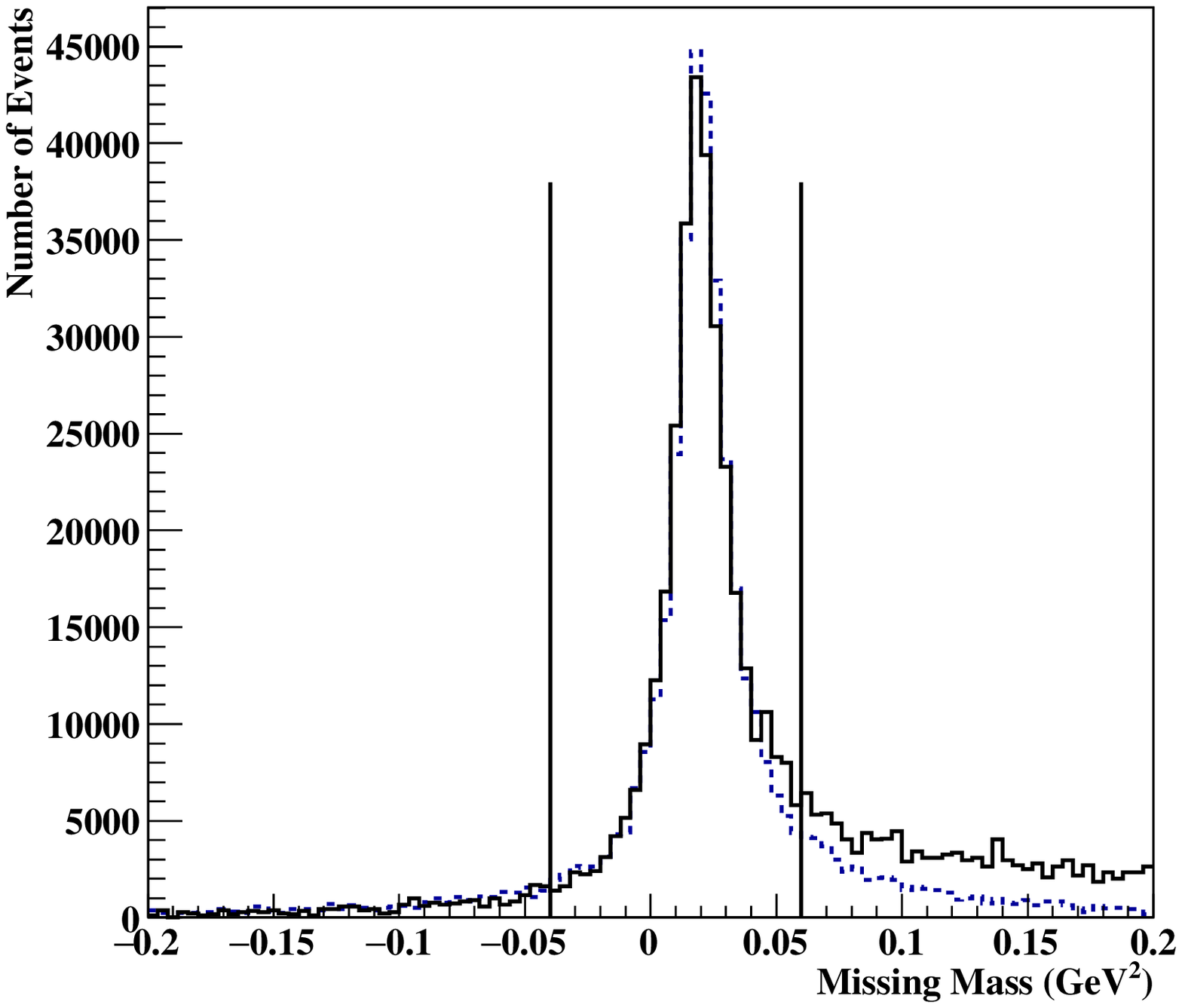}
\vspace{-0.1cm}
\caption{Square of the missing mass $M_{X}^2$ for $ep \to e'\pi^+\pi^-p'$, showing a peak 
at the $\pi^-$ mass squared. The dashed histogram is from the Monte Carlo and the solid 
histogram is the data. The vertical lines show the applied cut.}
\label{fig:missmass}
\end{center}
\end{figure*}

\begin{figure*}[htp]
\begin{center}
\includegraphics[width=13cm,keepaspectratio]{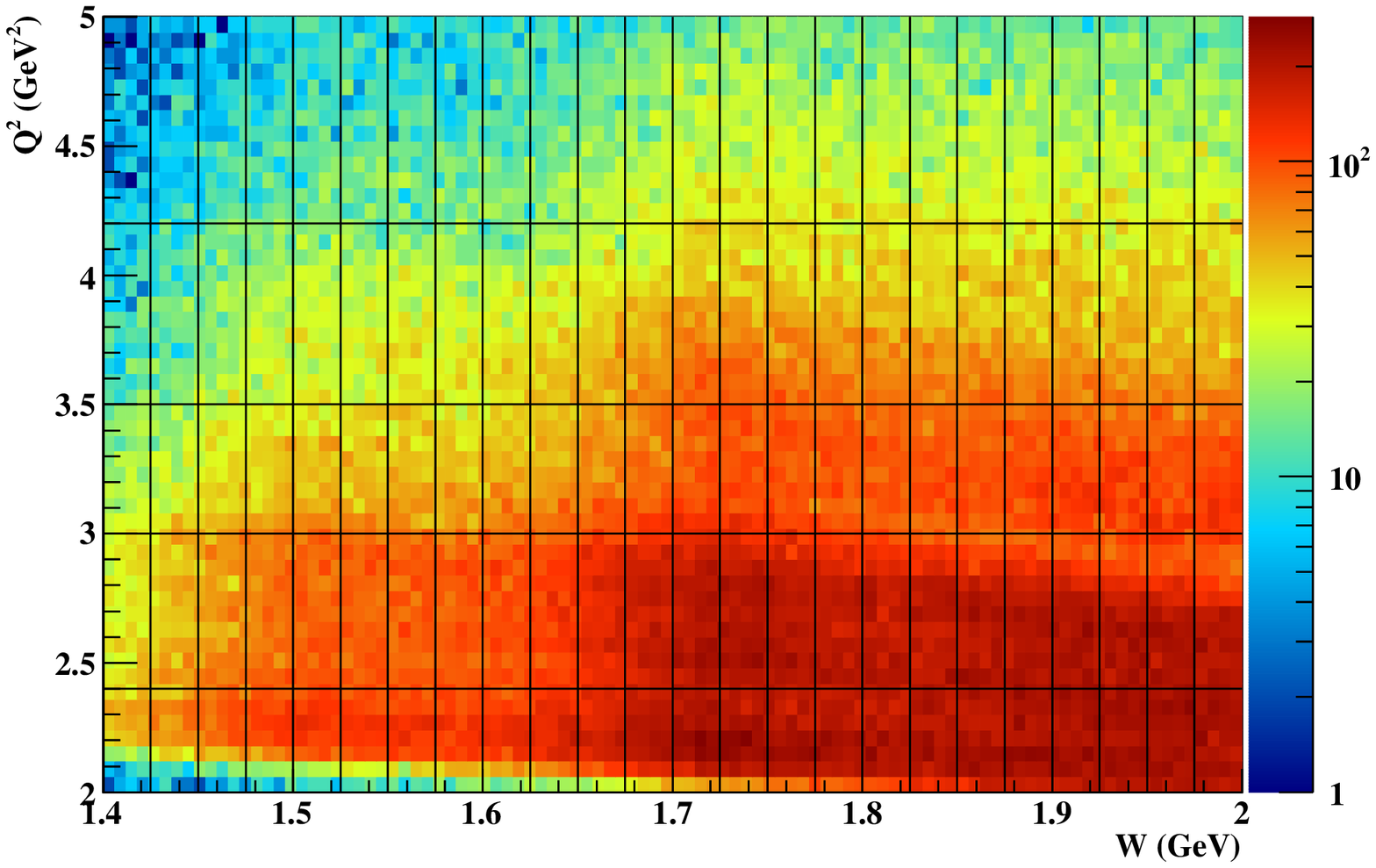}
\vspace{-0.1cm}
\caption{(Color Online) The kinematic coverage of the data, shown as a scatter-plot of 
events as a function of center of mass energy $W$ and squared 4-momentum transfer $Q^2$. 
Bins are shown within which the integrated and nine 1-fold differential $\pi^+\pi^-p$ 
cross sections were obtained.}
\label{fig:binning}
\end{center}
\end{figure*}

Events with a detected electron, proton, and positively charged pion were retained for 
further analysis. The reaction of interest here is $ep \to e'\pi^+\pi^-p'$, where the 
primed quantities are for the final state. The negative pion was bent toward the beamline 
and could bend outside of the detector acceptance. We reconstructed the mass of the 
$\pi^-$ using the missing mass technique.  The missing mass squared $M_{X}^{2}$ for these 
$ep \to e'p'\pi^+ X$ events is shown in Fig.~\ref{fig:missmass}, with a clean peak at the 
pion mass. The peak position and width compared very well with Monte Carlo (MC) simulated 
events. The larger number of events in the data at higher missing mass is due to radiative 
events, where the electron radiated a low-energy photon either before, after, or during the
scattering off the proton. The loss of these events from the peak was calculated using 
standard methods (described later in Section~\ref{radc}) and was corrected for in the final 
analysis. After all selections were applied, there remained 336,668 exclusive $\pi^+\pi^-p$ 
events. The distribution of data events for this measurement is shown in 
Fig.~\ref{fig:binning} as a function of the center of mass (CM) energy $W$ and the squared 
4-momentum transfer to the virtual photon $Q^2$. The data were binned, as shown by the black 
lines in the plot, to get the fully integrated cross section dependence on $W$ and $Q^2$.

\subsection{Reaction Kinematics}
\label{kin}

\begin{figure}[htbp]
\begin{center}
\includegraphics[width=8.25cm]{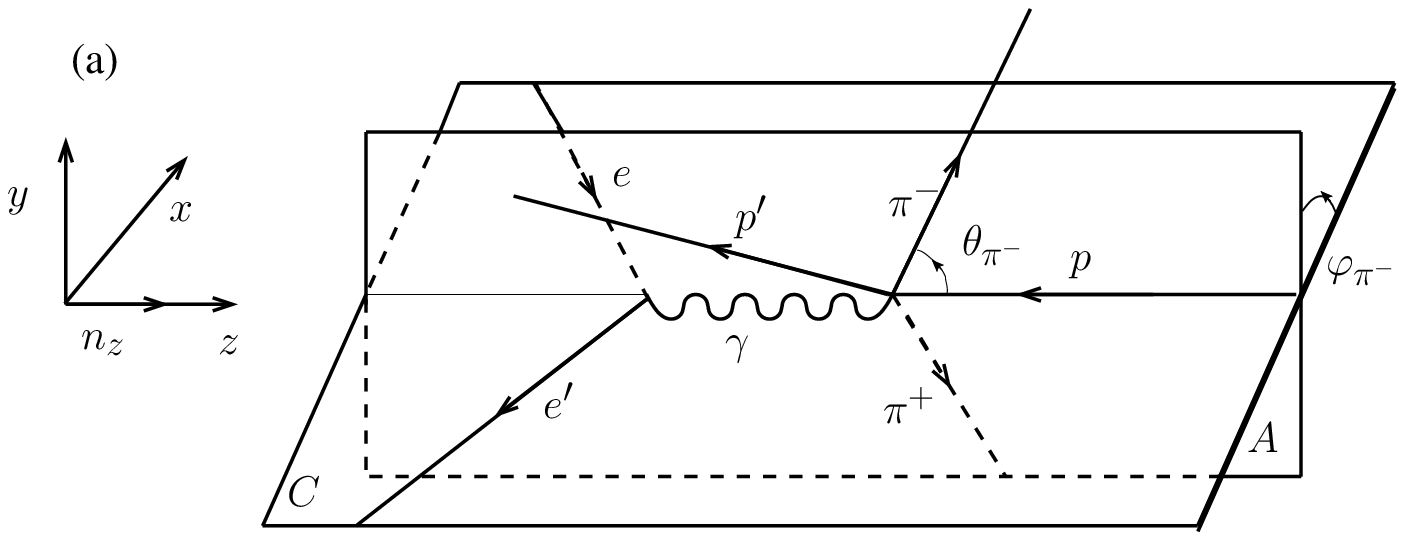}
\includegraphics[width=8.25cm]{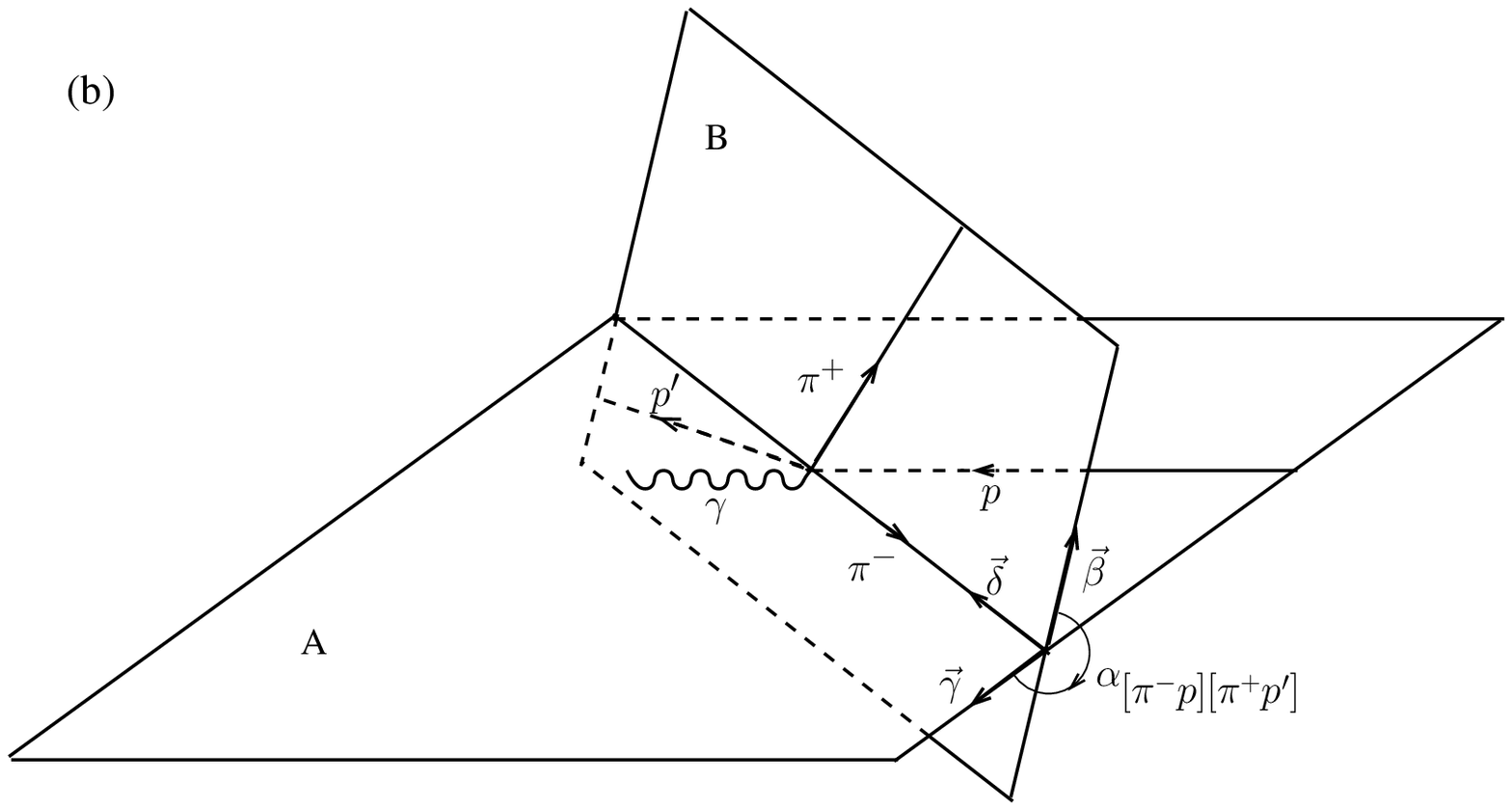}
\vspace{1cm}
\caption{Angular variables from the set defined by Eq.~(\ref{multibin}) 
for the description of the $e p \to e' \pi^+\pi^-p' $ reaction in the CM frame of 
the final state hadrons. Panel (a) shows the $\pi^-$ spherical angles $\theta_{\pi^-}$ and 
$\varphi_{\pi^-}$. Plane A is defined by the 3-momenta of the initial state proton and 
the final state $\pi^-$. Plane C represents the electron scattering plane. Panel 
(b) shows the angle $\alpha_{[\pi^-p][\pi^+p']}$  between the two defined hadronic planes 
A and B. Plane B is defined by the 3-momenta of the final state $\pi^+$ and $p'$. The unit 
vectors $\vec{\gamma}$ and $\vec{\beta}$ are normal to the $\pi^-$ 3-momentum in the 
planes A and B, respectively.} 
\label{kinematic}
\end{center}
\end{figure}

The kinematics of the reaction is shown in Fig.~\ref{kinematic}. The scattered electron 
defines a plane, which in our coordinate system is the $x-z$ plane. The direction of the 
$z$-axis was chosen to align with the virtual photon momentum vector. The $y$-axis is 
normal to the scattering plane with its direction defined by the vector product 
$\vec{n}_y = \vec{n}_z \times \vec{n}_x$ as shown in Fig.~\ref{kinematic}. The virtual 
photon and the outgoing $\pi^-$ form another plane, labeled A in Fig.~\ref{kinematic}, 
with angles $\theta$ and $\phi$ as shown. We also need the $\theta$ and $\phi$ angles for 
the $\pi^+$ and the final state proton $p'$, as described next.

Another plane is defined by the outgoing particles $\pi^+$ and $p'$, labeled B in 
Fig.~\ref{kinematic}, which intersects with plane A. Note that in the CM frame, the 
3-momenta of all three final state hadrons are located in the common plane B. The 
angle between the A and B planes is given by $\alpha_{[\pi^-p][\pi^+p']}$ as shown in 
Fig.~\ref{kinematic}. In order to calculate this angle, the vectors $\vec\beta$, 
$\vec\gamma$, and $\vec\delta$ are defined as shown in Fig.~\ref{kinematic} and 
evaluated as given in \cite{Fe09}. 


The 3-body final state is unambiguously determined by 5 kinematic variables. Indeed, 
three final state particles could be described by $4 \times 3 = 12$ components of their 
4-momenta. As each of these particles was on-shell, this provided three restrictions 
$E_i^2 -P_i^2 = m_i^2$~$(i=1,2,3)$. Energy-momentum conservation imposed four additional 
constraints for the final state particles, so that there were five remaining kinematic 
variables that unambiguously determine the 3-body final state kinematics. In the electron 
scattering process $e p \to e \pi^+\pi^-p'$, we also have the variables $W$ and $Q^2$ that 
fully define the initial state kinematics. So the electron scattering cross sections for 
double charged pion production should be 7-fold differential: 5 variables for the final 
state hadrons, plus $W$ and $Q^2$ determined by the electron scattering kinematics. Such 
7-fold differential cross sections may be written as $\frac{d^7\sigma}{dWdQ^2d^5\tau_i}$, 
where $d^5\tau$ is the 5-fold phase space for the final state hadron kinematics. Three 
sets of five kinematic variables were used with the spherical angles $\theta_i$ and 
$\varphi_i$ of the final state particle $\pi^-$, $\pi^+$, or $p'$, with the differentials 
labeled as $d^5\tau_i$, $i$= $\pi^-$, $\pi^+$, or $p'$, respectively. In addition to the 
spherical angles defined above, two other variables include the two invariant masses 
$M_{i,j}$ of the final state hadrons $i$ and $j$. The final variable represents the angle 
between the two planes A and B shown in Fig.~\ref{kinematic}, where plane A is formed by 
the three momenta of the initial state proton and the $i$-th final hadron, while plane B 
is formed by the pair of the three momenta of other two final state hadrons.

The five variables for $i=\pi^-$  $(M_{\pi^+\pi^-}, M_{\pi^+p'}$, $\theta_{\pi^-}$, 
$\varphi_{\pi^-}$, and $\alpha_{[\pi^-p][\pi^+p']})$  were calculated from the 3-momenta of 
the final state particles $\vec P_{\pi^-}$, $\vec P_{\pi^+}$, and $\vec P_{p'}$. Two other 
sets with respect to the $\pi^+$ and $p'$ were obtained by cyclic permutation of the 
aforementioned variables of the first set. All 3-momenta used from hereon, if not 
specified otherwise, were defined in the CM frame.


The $M_{\pi^+\pi^-}$ and $M_{\pi^+p'}$ invariant masses were related to the 4-momenta of 
the final state particles as
\begin{eqnarray}
\label{invmasses}
M_{\pi^+\pi^-} = \sqrt{(P_{\pi^+} + P_{\pi^-})^2} ~~\textrm{and} \nonumber \\
M_{\pi^+p'} = \sqrt{(P_{\pi^+} + P_{p'})^2} \textrm{ ,}
\end{eqnarray}
where $P_i$ represents the final state particle 4-momentum.

The angle $\theta_{\pi^-}$ between the 3-momentum of the initial state photon and the final 
state $\pi^-$ in the CM frame was calculated as
\begin{equation}
\theta_{\pi^-} = \cos^{-1}\left( \frac{(\vec P_{\pi^-} \cdot \vec P_\gamma)}
{|\vec P_{\pi^-}| |\vec P_\gamma|} \right).
\label{angletheta}
\end{equation}

The $\varphi_{\pi^-}$ angle was defined in a case-dependent manner by
\begin{eqnarray}
\varphi_{\pi^-} = \tan^{-1}\left( \frac{P_{y
\pi^-}}{P_{x\pi^-}} \right): &
P_{x\pi^-} > 0, P_{y\pi^-} > 0; \\
\varphi_{\pi^-} = \tan^{-1}\left( \frac{P_{y
\pi^-}}{P_{x\pi^-}} \right) + 2\pi: &
P_{x\pi^-} > 0, P_{y\pi^-} < 0; \\
\varphi_{\pi^-} = \tan^{-1}\left( \frac{P_{y
\pi^-}}{P_{x\pi^-}} \right) + \pi: &
P_{x\pi^-} < 0, P_{y\pi^-} < 0; \\
\varphi_{\pi^-} = \tan^{-1}\left( \frac{P_{y
\pi^-}}{P_{x\pi^-}} \right) + \pi: &
P_{x\pi^-} < 0 , P_{y\pi^-} > 0;  \\
\varphi_{\pi^-} = \pi/2: &
P_{x\pi^-} = 0, P_{y\pi^-} > 0;  \\
\varphi_{\pi^-} = 3\pi/2: &
P_{x\pi^-} = 0, P_{y\pi^-} < 0.
\end{eqnarray}

The calculation of the angle $\alpha_{[\pi^-p][\pi^+p']}$  between the planes A and 
B was more complicated. First we determined two auxiliary unit vectors $\vec \gamma$ 
and $\vec \beta$. The vector $\vec \gamma$ is perpendicular to the 3-momentum
$\vec P_{\pi^-}$, directed outward and situated in the plane given by the target proton 
3-momentum and the $\pi^-$ 3-momentum $\vec P_{\pi^-}$. The vector $\vec \beta$ is 
perpendicular to the 3-momentum of the $\pi^-$, directed toward the 
$\pi^+$ 3-momentum $\vec P_{\pi^+}$ and situated in the plane composed by the 
$\pi^+$ and $p'$ 3-momenta. As mentioned above, the 3-momenta of the $\pi^+$, $\pi^-$, 
and $p'$ were in the same plane, since in the CM frame their total 3-momentum must be 
equal to zero. The angle between the two planes A and B is then,
\begin{equation}
\alpha_{[\pi^-p][\pi^+p']} = \cos^{-1} (\vec \gamma \cdot \vec \beta),
\label{anglealpha}
\end{equation}
where the inverse cosine function runs between zero and $\pi$. On the other hand, the 
angle between the planes A and B may vary between zero and $2\pi$. To determine the 
angle $\alpha_{[\pi^-p][\pi^+p']}$  in a range between $\pi$ and $2\pi$, we looked at the 
relative direction of the vector $\vec P_{\pi^-}$ and the vector product of 
the unit vectors $\vec \gamma$ and $\vec \beta$,
\begin{equation}
\vec \delta = \vec \gamma \times \vec \beta.
\label{vecprod}
\end{equation}
If the vector $\vec \delta$ is collinear to $\vec P_{\pi^-}$, the 
$\alpha_{[\pi^-p][\pi^+p']}$ angle is determined by Eq.~(\ref{anglealpha}). In the case 
of anti-collinear vectors $\vec \delta$ and $\vec P_{\pi^-}$,
\begin{equation}
\alpha_{[\pi^-p][\pi^+p']} = 2\pi - \cos^{-1}(\vec \gamma \cdot \vec \beta).
\label{anglealpha_var}
\end{equation}
The vectors $\vec \gamma$, $\vec \beta$, and $\vec \delta$ may be expressed in terms of 
the final state hadron 3-momenta as given in \cite{Fe09}.

\subsection{Cross Section Formulation}
\label{crsec}

The 7-fold differential cross section may be written as 
\begin{equation}
\frac{d^7\sigma}{dWdQ^2dM_{\pi^+p'}dM_{\pi^+\pi^-}d\Omega_{\pi^-}
d\alpha_{[\pi^-p][[\pi^+p']}} \nonumber.
\label{crosssymb}
\end{equation} 
These cross sections were calculated from the quantity of selected events collected 
in the respective 7-dimensional cell as
\begin{equation}
\frac{d^7\sigma}{dWdQ^2d^5\tau} = 
\left( \frac{\Delta N}{{e\!f\!f} \cdot R} \right)
\left( \frac{1} {\Delta W \Delta Q^2 \Delta \tau_{\pi^-} L } \right),
\label{expcrossect}
\end{equation}
where $\Delta N$ is the number of events inside the 7-dimensional (7-d) bin,
$e\!f\!f$ is the efficiency for the $\pi^+\pi^-p$ event detection in the 7-d bin, 
$R$ is the radiative correction factor (described in Section~\ref{radc}), $L$ is 
the integrated luminosity (in units of $\mu b^{-1}$), $\Delta W$ and $\Delta Q^2$ 
are the binning in the electron scattering kinematics, and $\Delta \tau_{\pi^-}$ 
is the binning in the hadronic 5-d phase space:
\begin{equation}
\Delta \tau_{\pi^-} = 
\Delta M_{\pi^+p'} \Delta M_{\pi^+\pi^-} \Delta \cos(\theta_{\pi^-}) 
\Delta \varphi_{\pi^-} \Delta \alpha_{[\pi^-p][\pi^+p']} \ .
\label{multibin}
\end{equation}

In the one photon exchange approximation, the virtual photon cross section is related to
the electron scattering cross section by
\begin{equation}
\begin{aligned}
&\frac{d^5\sigma}{dM_{\pi^+p'}dM_{\pi^+\pi^-}d\Omega_{\pi^-} d\alpha_{[\pi^-p][\pi^+p']}} = \\
&\frac{1}{\Gamma_v}
\frac{d^7\sigma}{dWdQ^2dM_{\pi^+p'}dM_{\pi^+\pi^-}d\Omega_{\pi^-}
d\alpha_{[\pi-p][\pi^+p']}},
\label{fulldiff}
\end{aligned}
\end{equation}
where $\Gamma_v$ is the virtual photon flux given by
\begin{equation}
\Gamma_v = \frac{\alpha}{4\pi}\frac{1}{E_{beam}^2M_p^2}\frac{W(W^2-M_p^2)}
{(1-\varepsilon)Q^2},
\label{flux}
\end{equation}
and $\alpha$ is the fine structure constant, $M_p$ is the proton mass, and 
$\varepsilon$ is the virtual photon polarization parameter, 
\begin{equation}
\varepsilon = \left( 1 + 2\left( 1 + \frac{\omega^2}{Q^2} \right)
\tan^2\left(\frac{\theta_e}{2}\right) \right)^{-1}.
\label{polarization}
\end{equation}
Here $\omega = E_{beam} - E_{e'}$ and $\theta_e$ are the virtual photon energy and the 
electron polar angle in the lab frame, respectively, and $W$, $Q^2$, and $\theta_e$ are evaluated at 
the center of the bin.
The 7-d phase space for exclusive $ep \to e'\pi^+\pi^-p'$ electroproduction covered in our 
data set consists of 4,320,000 cells. Because of the correlation between the $\pi^+\pi^-$ 
and $\pi^+p'$ invariant masses of the final state hadrons imposed by energy-momentum 
conservation, only 3,606,120 7-d cells are kinematically allowed. They were populated by 
just 336,668 selected exclusive charged double pion electroproduction events. Most 7-d 
cells were either empty or contained just one measured event, which made it virtually 
impossible to evaluate the 7-fold differential electron scattering or 5-fold differential 
virtual photon cross sections from the data. Following previous studies~\cite{Fe09,Mo12,Ri03}, 
in order to achieve sufficient accuracy for these cross section measurements, the 5-fold 
differential cross sections were integrated over different sets of four variables, producing 
independent 1-fold differential cross sections. In the first step of physics analysis aimed 
at determining the contributing reaction mechanisms, it is even more beneficial to use the 
integrated 1-fold differential cross sections, since the structures and steep evolution of 
these cross sections elucidate the role of effective meson-baryon diagrams \cite{Mo09}. So in practice, 
we analyzed sets of 1-fold differential cross sections obtained by integration of the 5-fold 
differential cross sections over 4 variables in each bin of $W$ and $Q^2$. We used the 
following set of four 1-fold differential cross sections using $d^5\tau_{\pi^-}$ as expressed 
by Eq.~(\ref{multibin}):
\begin{eqnarray}
\label{inegr5diff}
\frac{d\sigma}{dM_{\pi^+\pi^-}} = &
\int\frac{d^5\sigma}{d^5\tau_{\pi^-}} 
dM_{\pi^+p'}d\Omega_{\pi^-}d\alpha_{[\pi^-p][\pi^+p']}, \nonumber \\
\frac{d\sigma}{dM_{\pi^+p'}} = &
\int\frac{d^5\sigma}{d^5\tau_{\pi^-}}
dM_{\pi^+\pi^-}d\Omega_{\pi^-}d\alpha_{[\pi^-p][\pi^+p']},  \\
\frac{d\sigma}{d(-\cos\theta_{\pi^-})} = &
\int\frac{d^5\sigma}{d^5\tau_{\pi^-}}
dM_{\pi^+\pi^-}dM_{\pi^+p'}d\varphi_{\pi^-}d\alpha_{[\pi^-p][\pi^+p']}, \nonumber \\
\frac{d\sigma}{d\alpha_{[\pi^-p][\pi^+p']}} = &
\int\frac{d^5\sigma}{d^5\tau_{\pi^-}}
dM_{\pi^+\pi^-}dM_{\pi^+p'}d\Omega_{\pi^-}. \nonumber 
\end{eqnarray}

Five other 1-fold differential cross sections were obtained by integration of the 5-fold 
differential cross sections defined over two different sets of kinematic variables 
with the $\pi^+$ and $p'$ solid angles, using $d^5 \tau_{\pi^+}$ and $d^5 \tau_{p'}$ 
defined analogously to Eq.~(\ref{multibin}):
\begin{eqnarray}
\label{inegr5diff1}
\frac{d\sigma}{d(-\cos\theta_{\pi^+})} = &
\int\frac{d^5\sigma}{d^5\tau_{\pi^+}}
dM_{\pi^-p'}dM_{\pi^+p'}d\varphi_{\pi^+}d\alpha_{[\pi^+p][\pi^-p']}, \nonumber \\
\frac{d\sigma}{d\alpha_{[\pi^+p][\pi^-p']}} = &
\int\frac{d^5\sigma}{d^5\tau_{\pi^+}}
dM_{\pi^-p'}dM_{\pi^+p'}d\Omega_{\pi^+},   \\
\frac{d\sigma}{dM_{\pi^-p'}} = &
\int\frac{d^5\sigma}{d^5\tau_{\pi^+}}
dM_{\pi^+p'}d\Omega_{\pi^+}d\alpha_{[\pi^+p][\pi^-p']},\nonumber \\
\frac{d\sigma}{d(-\cos\theta_{p'})} = &
\int\frac{d^5\sigma}{d^5\tau_{p'}}
dM_{\pi^+\pi^-}dM_{\pi^-p'}d\varphi_{p'}d\alpha_{[p'p][\pi^+\pi^-]}, \nonumber \\
\frac{d\sigma}{d\alpha_{[p'p][\pi^+\pi^-]}} = &
\int\frac{d^5\sigma}{d^5\tau_{p'}}
dM_{\pi^+\pi^-}dM_{\pi^-p'}d\Omega_{p'}. \nonumber
\end{eqnarray}

The statistical uncertainties for the 1-fold differential cross sections obtained from 
our data are in the range from 14\% at the smallest photon virtuality ($Q^2$=2.1~GeV$^2$) 
to 20\% at the biggest photon virtuality ($Q^2$=4.6~GeV$^2$), which are comparable with 
the uncertainties achieved with our previous CLAS data~\cite{Fe09,Ri03} from which resonance 
electrocouplings were successfully extracted~\cite{Mo16,Mo12}. 

\subsection{Detector Simulations and Efficiencies}
\label{mc}

The Monte Carlo event generator employed for the acceptance studies was similar to that 
described in \cite{Gol12}. This event generator is capable of simulating the event 
distribution for the major meson photo- and electroproduction channels in the $N^*$ 
excitation region. The input to the event generator included various kinematical parameters 
($W$, $Q^2$, electron angles, and so on) along with a description of the hydrogen target 
geometry. This event generator also included radiative effects, calculated according to 
\cite{Mo69}. Simulation of $\pi^+\pi^-p$ electroproduction events was based on the 
old version of the JLab-MSU model JM06~\cite{Mo06,Mo01,Ri00}, adjusted to reproduce the 
measured event kinematic distributions. The generated events were fed into the standard 
CLAS detector simulation software, based on CERN's GEANT package, called GSIM. The detector 
efficiency for a given 7-d kinematic bin was given by
\begin{equation}
e\!f\!f = \frac{N_{rec}}{N_{gen}},
\label{acceptance}
\end{equation}
where $N_{gen}$ is the number of events generated for a given kinematic bin and $N_{rec}$ 
the number of events reconstructed by the GSIM software. The same detector fiducial 
volume was used for both data and simulations to restrict the reconstructed tracks to the 
regions of the CLAS detector where efficiency evaluations were reliable. After applying the 
fiducial cuts, the detector efficiency tables for a given kinematic bin were determined in 
order to be used to calculate the cross sections.

In the data analysis for some 7-d cells, there was a reasonable number (more than 10) of 
generated simulation events, but the quantity of accepted events was equal to zero. Such 
situations represent an indication of zero CLAS detector acceptance in these kinematic 
regions. It was necessary to account for the contribution of such ``blind" areas to the 
integrals for the 1-fold differential cross sections given above.

\begin{figure*}[htp]
\begin{center}
\includegraphics[width=13cm,keepaspectratio]{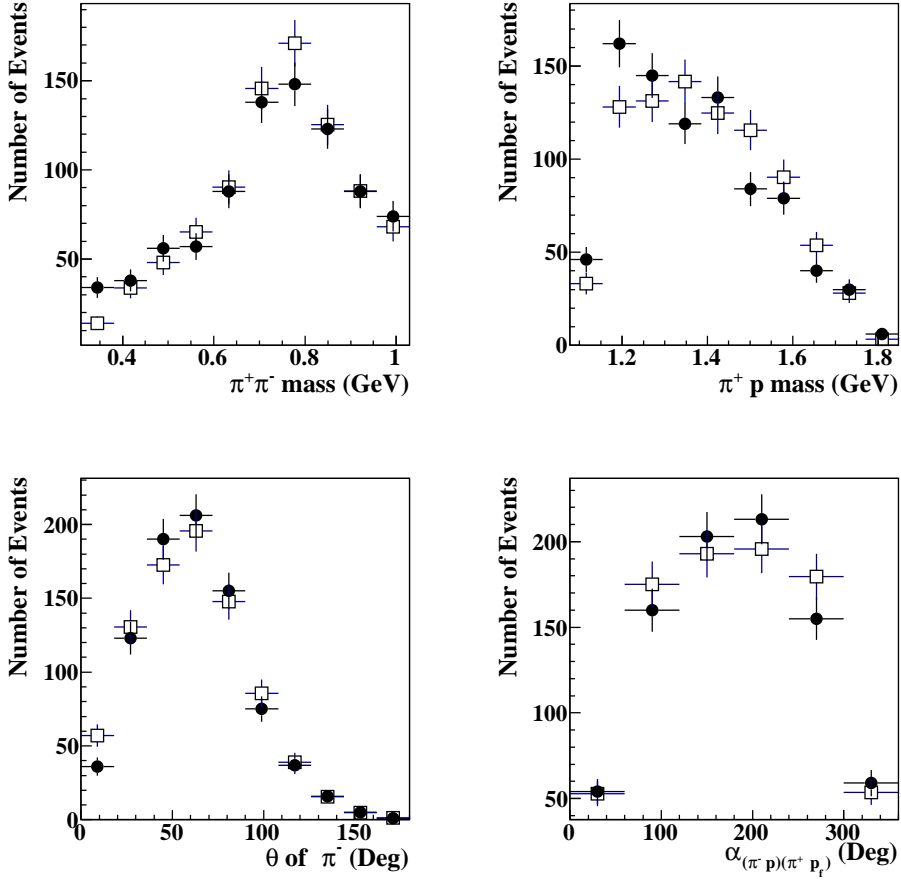}
\vspace{-0.1cm}
\caption{A comparison between the measured event distributions (solid circles) and the 
simulated event distributions (open squares) within the framework of the JM06 model
\cite{Ri00,Mo01,Mo06}, which was further adjusted in order to reproduce the measured 
event distributions. These comparisons are shown for the bin of $W$=1.99~GeV and 
$Q^2$=4.6~GeV$^2$.}
\label{fig:egmessim}
\end{center}
\end{figure*}

To estimate the contributions to the cross sections from detector blind areas, we used 
information from the event generator. We evaluated such contributions based on the cross 
section description of the JM06 event generator. The JM06 model~\cite{Mo06,Mo01,Ri00} was 
not previously compared with charged double pion electroproduction data at $Q^2 > 2.0$~GeV$^2$. 
Therefore, the JM06 model was further adjusted to the measured event distributions over the  
$\pi^+\pi^-p$ final state kinematic variables discussed above. After adjustment, the event 
generator gave a fair description of the data on the measured event distributions over the 
kinematic variables for all 1-fold differential cross sections. As a representative example, 
a comparison between the measured and simulated event distributions is shown in 
Fig.~\ref{fig:egmessim}. A comparable quality of agreement was achieved over the entire 
kinematic range covered by our measurements.

To obtain the 5-fold differential virtual photon cross sections in the blind areas
we used: 
\begin{itemize}
\item the number of measured data events (we weighted these events with the integral 
efficiency inside the 5-d bin) in the current $(W,Q^2)$ bin, integrated over all hadronic 
variables for the $\pi^+\pi^-p$ final state $N_{data,int}$;

\item the number of these events estimated from the event generator $N_{generated,int}$; and

\item the number of generated events in a 7-d blind kinematic bin $(W,Q^2,\tau_i)$, which
we call $N_{generated}^{7d}$. 
\end{itemize}

Using the event generator as a guide, we interpolated the number of events measured 
outside of the blind bin into the blind bin. Thus, the number of counts for the 7-fold
differential cross sections {\it in the blind bins only} were calculated by
\begin{equation}
\Delta N = \frac{N_{data,int}}{N_{generated,int}}N_{generated}^{7d},
\label{emptycell}
\end{equation}
and the 5-fold differential virtual photon cross sections in the blind bins were computed 
from $\Delta N$ in according to Eqs.~(\ref{expcrossect}-\ref{polarization}),
where we set $e\!f\!f=1$.

\begin{figure*}[htp]
\begin{center}
\includegraphics[width=13cm,keepaspectratio]{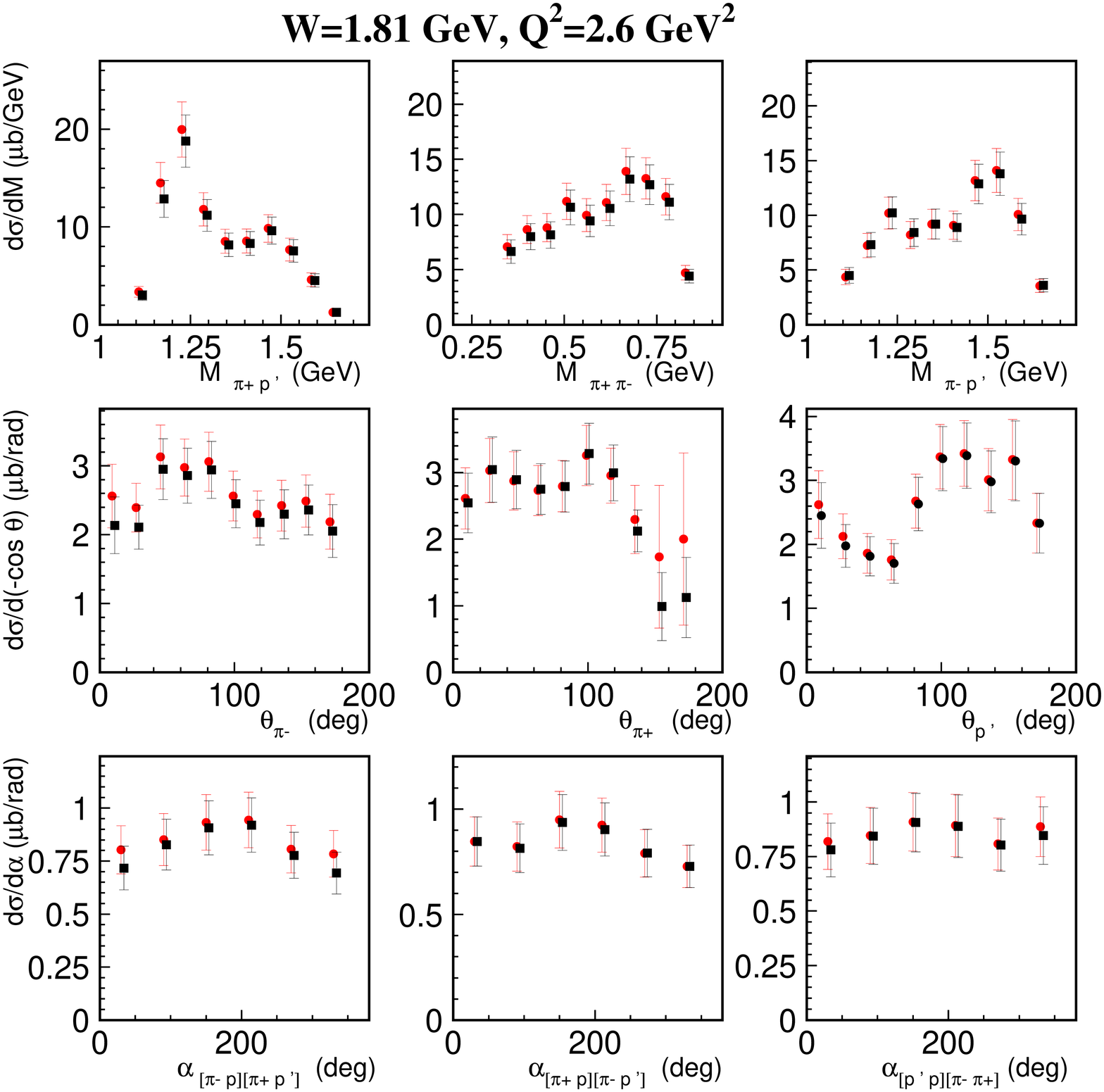}
\vspace{-0.1cm}
\caption{(Color online) Impact of the interpolation of the 5-fold $\pi^+\pi^-p$ differential 
cross sections into the blind areas of CLAS to the nine 1-fold differential cross sections at 
$W$=1.81~GeV and $Q^2$=2.6~GeV$^2$. The 1-fold differential cross sections obtained assuming 
zero 5-fold differential cross sections and the interpolated values for these cross sections 
in the blind areas of CLAS are shown by the black squares and red circles, respectively. 
The error bars represent statistical uncertainties.
To aid visualization, we have slightly shifted horizontally the two data sets.}
\label{fig:comp_fullempty}
\end{center}
\end{figure*}

A comparison between the 1-fold differential cross sections obtained with and without 
generated events inside the blind bins is shown in Fig.~\ref{fig:comp_fullempty}. Except 
for the two bins of maximal CM $\theta_{\pi^+}$ angles, the difference between the two 
methods is rather small, and is inside the statistical uncertainties for most points. The 
estimated uncertainty introduced by this interpolation method has an upper limit of 5\% on 
average, depending on the kinematics.  

\subsection{Radiative Corrections}
\label{radc}

\begin{figure*}[htp]
\begin{center}
\includegraphics[width=13cm,keepaspectratio]{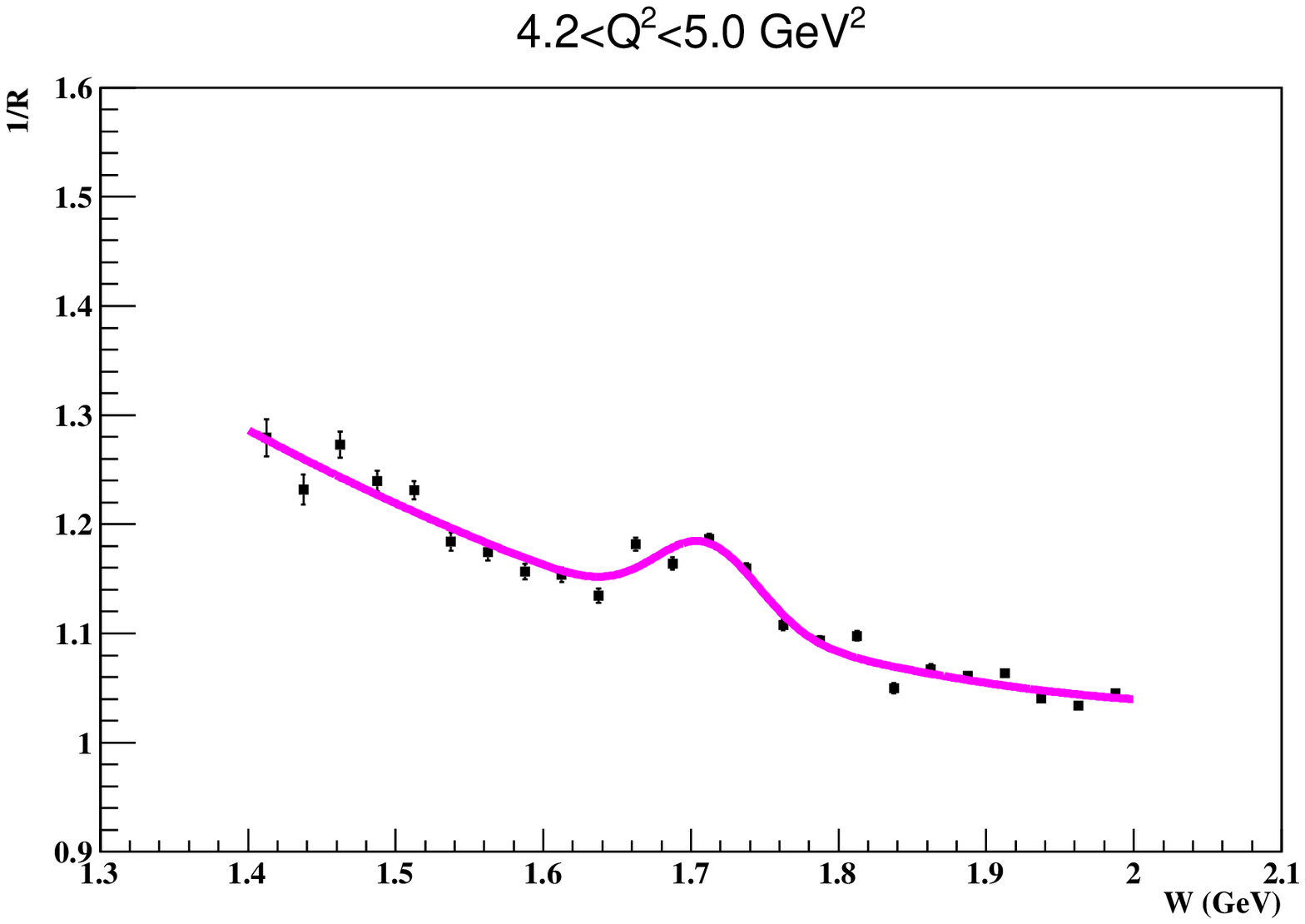}
\vspace{-0.1cm}
\caption{(Color Online) The radiative correction factor $1/R$ for the bin 
4.2~GeV$^2$ $< Q^2 <5.0$~GeV$^2$. The solid magenta line represents a polynomial plus
Gaussian fit.}
\label{fig:radc}
\end{center}
\end{figure*}

To estimate the influence of radiative correction effects, we simulated 
$ep \to e'\pi^+\pi^-p'$ events using the above event generator both with and without 
radiative effects. For the simulation of radiative effects in double pion 
electroproduction, the well known Mo and Tsai procedure~\cite{Mo69} was used. As 
described above, we integrated the 5-fold two pion cross sections over four variables to 
get 1-fold differential cross sections. This integration considerably reduced the 
influence of the final state hadron kinematic variables on the radiative correction 
factors for the analyzed 1-fold differential cross sections. The radiative correction 
factor $R$ in Eq.~(\ref{expcrossect}) was determined as
\begin{equation}
\label{radcorrfact}
R = \frac{N_{rad}^{2d}}{N_{norad}^{2d}},
\end{equation}
where $N_{rad}^{2d}$ and $N_{norad}^{2d}$ are the numbers of generated events in each 
$(W,Q^2)$ bin with and without radiative effects, respectively. We then fit the inverse 
factor $1/R$ over the $W$ range in each $Q^2$ bin. The factor $1/R$ for a representative 
bin 4.2~GeV$^2$ $< Q^2 < 5.0$~GeV$^2$ is plotted as a function of $W$ in Fig.~\ref{fig:radc}. 
A few words should be said about the behavior of this factor. Since the radiation migrates 
events from lower $W$ to higher $W$, and because the structure at $W$ of around $1.7$~GeV 
is the most prominent feature of the cross sections, there is a small enhancing bump in the 
factor $1/R$ present in each $Q^2$ bin. 

\subsection{Systematic Uncertainties}
\label{systuncer}

\begin{figure*}[htp]
\begin{center}
\includegraphics[width=13cm,keepaspectratio]{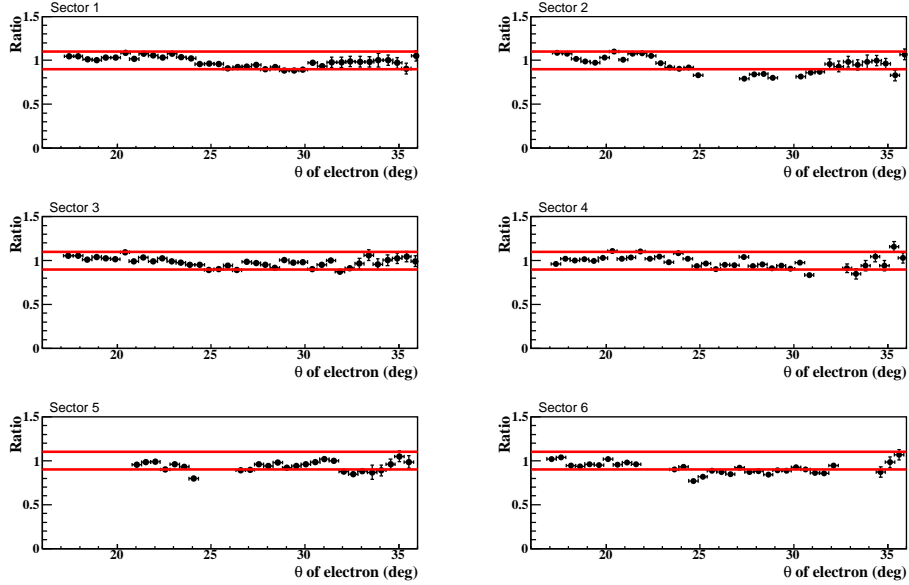}
\vspace{-0.1cm}
\caption{(Color Online) Ratio of the elastic cross section to the Bosted parameterization
\cite{Bost} as a function of electron polar angle $\theta$ for each of the six sectors of 
CLAS. The regions where there are missing data are the result of $\theta$ vs. $p$ cuts
to remove problematic areas of the detector. The horizontal lines represent $\pm$10\% 
deviations of the ratio from unity.}
\label{fig:elast_bosted}
\end{center}
\end{figure*}

One of the main sources of systematic uncertainty in this experiment is the uncertainty 
in the yield normalization factors, including the acceptance corrections, electron
identification efficiency, detector efficiencies, and beam-target luminosity. The elastic 
events present in the data set were used to check the normalization 
of the cross sections by comparing the measured elastic cross sections to the world data. 
This allowed us to combine the luminosity normalization, electron detection, electron tracking, and 
electron identification uncertainties into one global uncertainty factor. In 
Fig.~\ref{fig:elast_bosted} the ratio of the elastic cross section to the Bosted 
parameterization~\cite{Bost} is shown. The parameterized cross section and that from the 
CLAS elastic data are shown after accounting for radiative effects so that they are
directly comparable. One can see most of the points are positioned within the red lines 
that indicate $\pm$10\% offsets. This comparison allowed us to assign a conservative 10\% 
point-to-point uncertainty to the full set of yield normalization factors for the two pion 
cross sections.

We restricted the $ep \to e'\pi^+p'X$ missing mass to be close to the $\pi^-$ peak in order 
to select two pion events. This missing mass cut event selection caused some loss of events. 
Uncertainties due to such losses were estimated  by using Monte Carlo simulations for the 
acceptance calculations. The initial Monte Carlo distributions had better resolution than 
the data, so special CLAS software (GPP) was used to make them match. The uncertainty 
associated with the missing mass cuts was estimated by calculating the difference in the 
cross sections with two different missing mass cuts applied both on the real data and the
Monte Carlo data sample. The missing mass cut used in the analysis was 
-0.04~GeV$^2$ $< M_{\pi^-X}^2 <0.06$~GeV$^2$, so we varied the range of this cut to 
-0.02~GeV$^2$ $< M_{\pi^-X}^2 <0.03$~GeV$^2$ to estimate the systematic uncertainty due to 
the missing mass cut. 

We used the following method for estimating systematic uncertainties. In each case for a 
given observable ($e.g.$, mass distributions) we calculated the relative difference 
$(\sigma-\sigma_c)/\sigma$, where $\sigma_c$ is the recalculated cross section with a more 
narrow missing mass cut. We expected to see a Gaussian-like distribution for the relative 
difference distribution. The difference between the centroid of this distribution and zero 
is a measure of the systematic uncertainty. From this, we estimated the systematic 
uncertainty due to the missing mass cuts at about 4.2\% of the measured differential cross 
sections. 

To estimate the influence of the detector fiducial area cuts, we recalculated the cross 
sections without applying fiducial cuts to the hadrons. Again, we constructed the relative 
difference $(\sigma-\sigma_c)/\sigma$, where $\sigma_c$ is the recalculated cross section 
without hadron fiducial cuts. The result is that we saw a systematic decrease of about 
2\% in the cross sections.  

We also varied the particle identification criteria, which included a cut on the calculated 
speed and momentum of the detected hadrons. In our analysis we applied a $\pm$2$\sigma$ 
cut, so to estimate the influence of these cuts to our results we recalculated cross 
sections with a $\pm$3$\sigma$ cut. By widening the particle identification cuts and 
using the same relative difference procedure as above, we saw a systematic increase of 
about 4.6\% of the cross sections. 

In addition, there were additional point-to-point uncertainties, dependent on the 5-d 
kinematics, due to the interpolation procedure to fill the blind bins. This systematic 
uncertainty for the 1-fold differential cross sections was estimated (from the differences 
shown in Fig.~\ref{fig:comp_fullempty}) to be on average 5\% as an upper limit, but may be 
smaller in regions where the JM06 model gave a good representation of the measured cross 
sections and where we have only small contributions from filling blind areas of CLAS.
Adding in quadrature the various systematic uncertainties, which were dominated by the 
normalization corrections, we found an overall systematic uncertainty of 14\% 
for the cross sections reported here. The summary of the systematic uncertainties can 
be found in Table~\ref{tab:systtab}. 

\begin{table}[h!]
\vspace{0.5cm}
\begin{tabular}{|l|d{4.3}|} \hline
\multicolumn{1} {|c|}{Sources of systematics} & \multicolumn{1} {|c|}{uncertainty,~$\%$} \\ \hline
Yield normalization   & 10.0 \\ \hline
Missing mass cut      &  4.2 \\ \hline
Hadron fiducial cuts  &  2.0 \\ \hline
Hadron ID cuts        &  4.6 \\ \hline
Radiative corrections &  5.0 \\ \hline
Event generator       &  5.0 \\  \hline 
Total                 & 14.0 \\ \hline
\end{tabular}
\caption{Summary of sources of point-to-point systematic uncertainties for the cross section
measurements reported in this work.}
\label{tab:systtab}
\end{table}

\section{Results and Discussion}
\label{impact}

\begin{figure*}[htp]
\begin{center}
\includegraphics[width=12.cm]{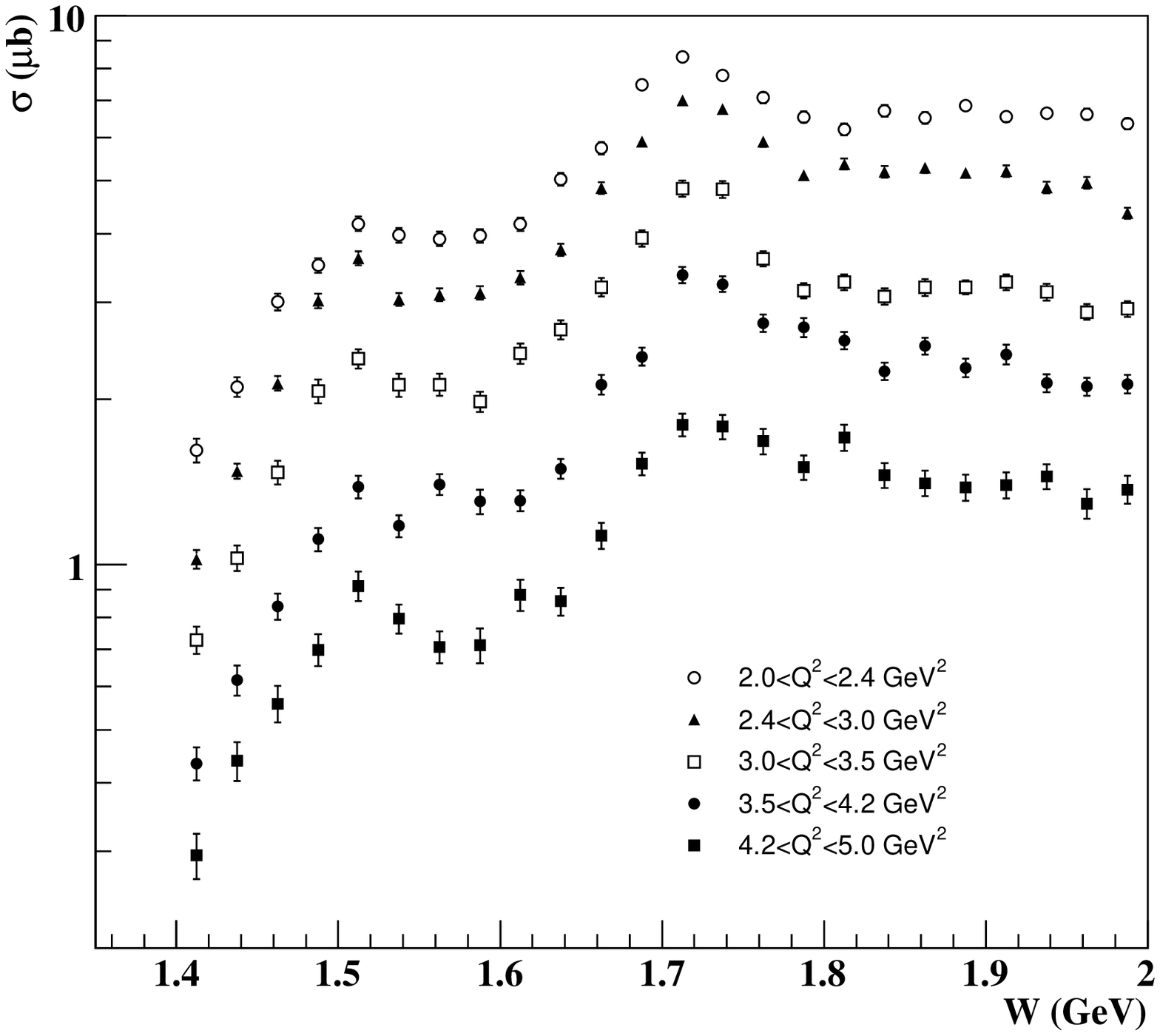}
\vspace{-0.1cm}
\caption{Fully integrated cross sections for $\pi^+\pi^-p$ electroproduction off protons at 
photon virtualities $Q^2$=2.2, 2.6, 3.2, 3.8, 4.6~GeV$^2$. The error bars represent the 
statistical uncertainties.}
\label{total_sec}
\end{center}
\end{figure*}

The fully integrated $\pi^+\pi^-p$ electroproduction cross sections obtained by integration 
of the 5-fold differential cross sections are shown in Fig.~\ref{total_sec} for five $Q^2$
bins. Two structures located at $W$=1.5~GeV and 1.7~GeV produced by the resonances of the 
second and third resonance regions are the major features in the $W$ evolution of the 
integrated cross sections observed in the entire range of $Q^2$ covered by the CLAS 
measurements. 

The results on the $\pi^+\pi^-p$ electroproduction cross sections discussed in 
Section~\ref{expt} open up the possibility to extend our knowledge of the $\gamma_vpN^*$ 
electrocouplings of many resonances up to photon virtualities $Q^2=5$~GeV$^2$, in particular 
for the states in the mass range above 1.6~GeV~\cite{Mo16,Mo16a}, which decay 
preferentially to $\pi\pi N$ final states. This $Q^2$ range corresponds 
to the distance scale where the transition to the dominance of quark core contributions to the 
resonance structure takes place~\cite{Bu12,Az13,Seg14,Seg15}. 

Here, we discuss the prospects for the extraction of resonance parameters from the new data 
based on comparisons between the measured nine 1-fold differential cross sections and the 
projected resonant contributions. Resonant contributions are computed within the framework 
of the recent JM model version~\cite{Mo09,Mo12,Mo16} employing the unitarized Breit-Wigner 
ansatz for the resonant amplitudes described in \cite{Mo12} and using interpolated  
resonance electrocouplings previously extracted in the analyses of exclusive meson 
electroproduction data from CLAS~\cite{Bu12,Az13,Park15}. This new version of the JM model
is here referred to as JM16.

\begin{table*}
\begin{center}
\begin{tabular}{|c|c|c|} \hline
Exclusive meson            &  Nucleon                        & $Q^2$ ranges for extracted \\
electroproduction channels &  resonances                     & $\gamma_vpN^*$ electrocouplings,~GeV$^2$ \\ \hline
$\pi^0p$, $\pi^+n$         & $\Delta(1232)3/2^+$,                              & 0.16-6.00 \\
                           & $N(1440)1/2^+$, $N(1520)3/2^-$, $N(1535)1/2^-$    & 0.30-4.16 \\ \hline
$\pi^+n$                   & $N(1675)5/2^-$, $N(1680)5/2^+$                    & 1.6-4.5 \\
                           & $N(1710)1/2^+$                                    & 1.6-4.5 \\ \hline
$\eta p$                   & $N(1535)1/2^-$                                    & 0.2-2.9 \\ \hline
$\pi^+ \pi^- p$            & $N(1440)1/2^+$, $N(1520)3/2^-$                    & 0.25-1.50 \\ 
                           & $\Delta(1620)1/2^-$, $N(1650)1/2^-$, $N(1680)5/2^+$  & 0.50-1.50 \\
                           & $\Delta(1700)3/2^-$, $N(1720)3/2^+$, $N'(1720)3/2^+$  & 0.50-1.50 \\  \hline 
\end{tabular}
\caption{Summary of the results on the nucleon resonance electrocouplings 
available from analyses of the CLAS exclusive meson electroproduction data 
off protons \cite{Mo16,Bu12,Az09,Park15,Mo12,Mo14}.}
\label{reselcoupl} 
\end{center}
\end{table*}

\begin{table*}
\begin{center}
\begin{tabular}{|c|c|c|d{4.3}|} \hline
Resonances      &$\Gamma_{tot},$ & Branching fraction & \multicolumn{1}{|c|}{Branching fraction} \\
                & MeV           &  to $\pi\Delta$, \% & \multicolumn{1}{|c|}{to $\rho p$, \%} \\ \hline
$N(1440)1/2^+$      & 387    & 19     & 1.7  \\
$N(1520)3/2^-$      & 130    & 25     & 9.4  \\
$N(1535)1/2^-$      & 131    & 2      & 10   \\
$\Delta(1620)1/2^-$ & 158    & 43     & 49   \\
$N(1650)1/2^-$      & 155    & 5      & 6    \\ 
$N(1680)5/2^+$      & 115    & 21     & 13   \\
$\Delta(1700)3/2^-$ & 276    & 84     & 5    \\
$N(1700)3/2^-$      & 148    & 45     & 52   \\
$N'(1720)3/2^+$     & 115    & 51     & 9    \\
$N(1720)3/2^+$      & 117    & 39     & 44   \\ \hline
\end{tabular}
\caption{The nucleon resonances included in the evaluation of the resonant contributions 
to the $\pi^+\pi^-p$ electroproduction cross sections off protons, and their total decay 
widths and branching fractions for decays to the $\pi\Delta$ and $\rho p$ final hadron 
states used in the evaluation of the resonant contributions to the current measurements.}
\label{hadrdec}  
\end{center}
\end{table*}

So far, $\gamma_vpN^*$ electrocouplings are available for excited nucleon states in the 
mass range up to 1.8~GeV. They were obtained from various CLAS data in the exclusive 
channels: $\pi^+n$ and $\pi^0p$ at $Q^2 < 5.0$~GeV$^2$ in the mass range up to 
1.7~GeV, $\eta p$ at $Q^2 < 4.0$~GeV$^2$ in the mass range up to 1.6~GeV, and 
$\pi^+\pi^-p$ at $Q^2 < 1.5$~GeV$^2$ in the mass range up to 1.8~GeV. A summary of the 
results on the available resonance $\gamma_vpN^*$ electrocouplings can be found in 
Table~\ref{reselcoupl}. The $\gamma_vpN^*$ electrocoupling values, together with the 
appropriate references, are available from our web page~\cite{resnum}.

The $\gamma_vpN^*$ electrocouplings employed in the evaluations of the resonant 
contributions to the $\pi^+\pi^-p$ differential cross sections were obtained from 
interpolation or extrapolation of the experimental results~\cite{resnum} by polynomial 
functions of $Q^2$. The estimated resonance electrocouplings can be found in \cite{Is16}. 
For low-lying excited nucleon states in the mass range $M_{N^*} <$ 1.6~GeV, the 
experimental results on the $\gamma_vpN^*$ electrocouplings are available at photon 
virtualities up to 5.0~GeV$^2$. Electrocouplings of these resonances were estimated by 
interpolating the data points. Electrocouplings of the $N(1675)5/2^-$, $N(1680)5/2^+$, and 
$N(1710)1/2^+$ resonances are available from $\pi^+n$ electroproduction data~\cite{Park15} 
at $Q^2$ from 2.0~GeV$^2$ to 5.0~GeV$^2$. To estimate their contributions to the 
$\pi^+\pi^-p$ electroproduction cross sections, we interpolated those results in $Q^2$. 

Electrocouplings of the $\Delta(1620)1/2^-$, $\Delta(1700)3/2^-$, and $N(1720)3/2^+$ 
resonances are available at $Q^2 < 1.5$~GeV$^2$~\cite{Mo14,Mo16,Mo16a}. The recent 
combined analysis of the CLAS $\pi^+\pi^-p$ electroproduction off proton data~\cite{Ri03} 
and the preliminary $\pi^+\pi^-p$ photoproduction data have revealed a contribution from 
a new candidate $N'(1720)3/2^+$ state~\cite{Mo16a}. This new $N'(1720)3/2^+$ state and 
the existing $N(1720)3/2^+$ state with very similar masses and total hadronic decay widths, 
have distinctively different hadronic decays to the $\Delta \pi$ and $N\rho$ final states,
and a very different $Q^2$-evolution of their associated electrocouplings. The resonant 
part of the $\pi^+\pi^-p$ electroproduction cross sections was computed by extrapolating 
the available results to the range of photon virtualities 2.0~GeV$^2$ $< Q^2 < 5.0$~GeV$^2$.

The contributions from resonances in the mass range above 1.8~GeV were not taken into 
account due to the lack of experimental results on their electrocouplings, thus 
limiting our evaluation of the resonant contributions to the range of $W<1.8$~GeV. 

The hadronic decay widths to the $\pi\Delta$ and $\rho p$ final states for the above 
resonances were taken from previous analyses of the CLAS $\pi^+\pi^-p$ electroproduction 
data off protons~\cite{Mo12,Mo14,Mo16,Mo16a}. The constraints imposed by the requirement 
to describe $\pi^+\pi^-p$ electroproduction data with $Q^2$ independent hadronic decay 
widths for the contributing states, allowed us to obtain improved estimates of the branching 
fractions (BF) for the resonances listed in Table~\ref{hadrdec}. 

\begin{figure*}[htp]
\begin{center}
\includegraphics[width=5.5cm]{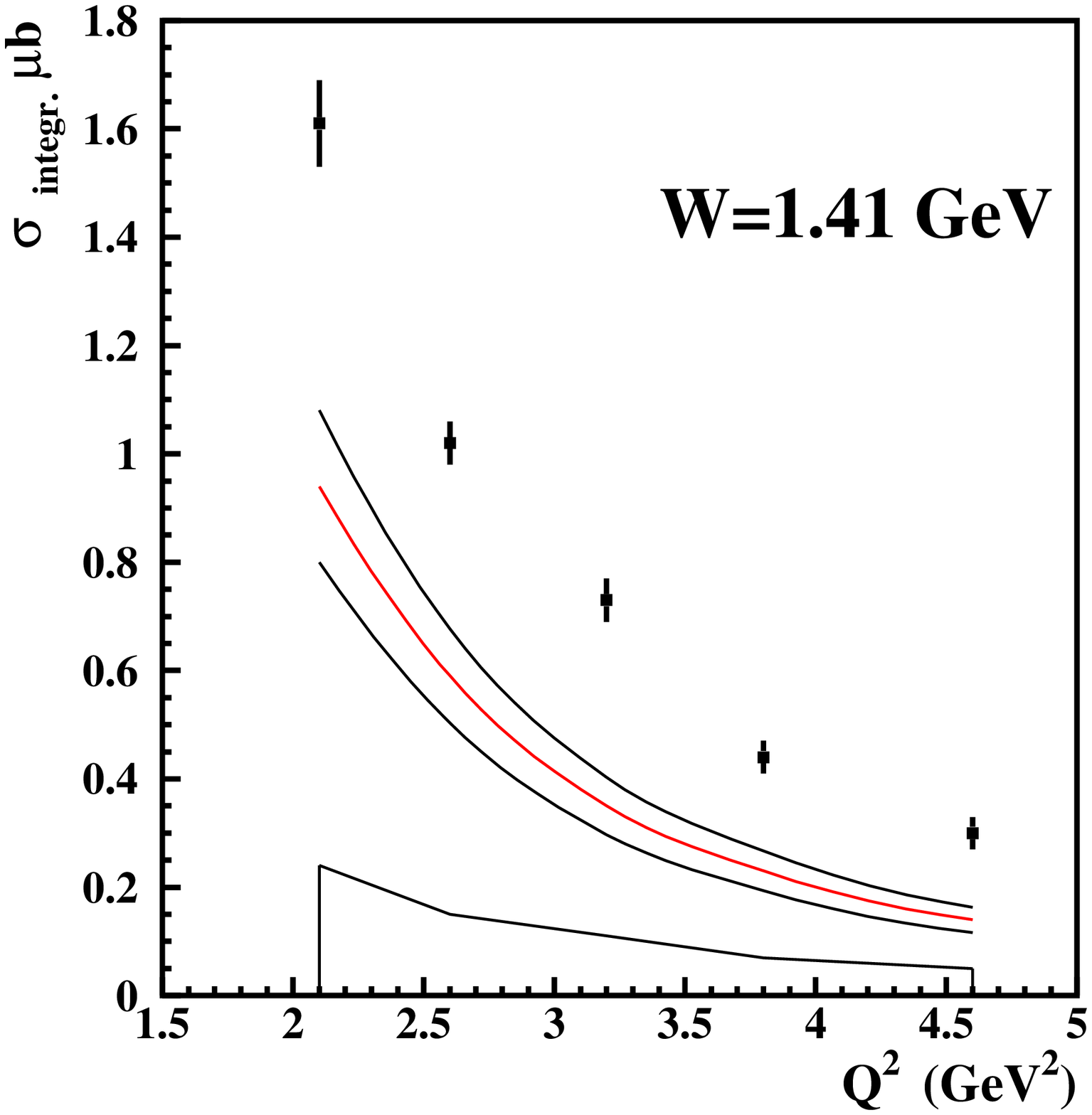}
\includegraphics[width=5.5cm]{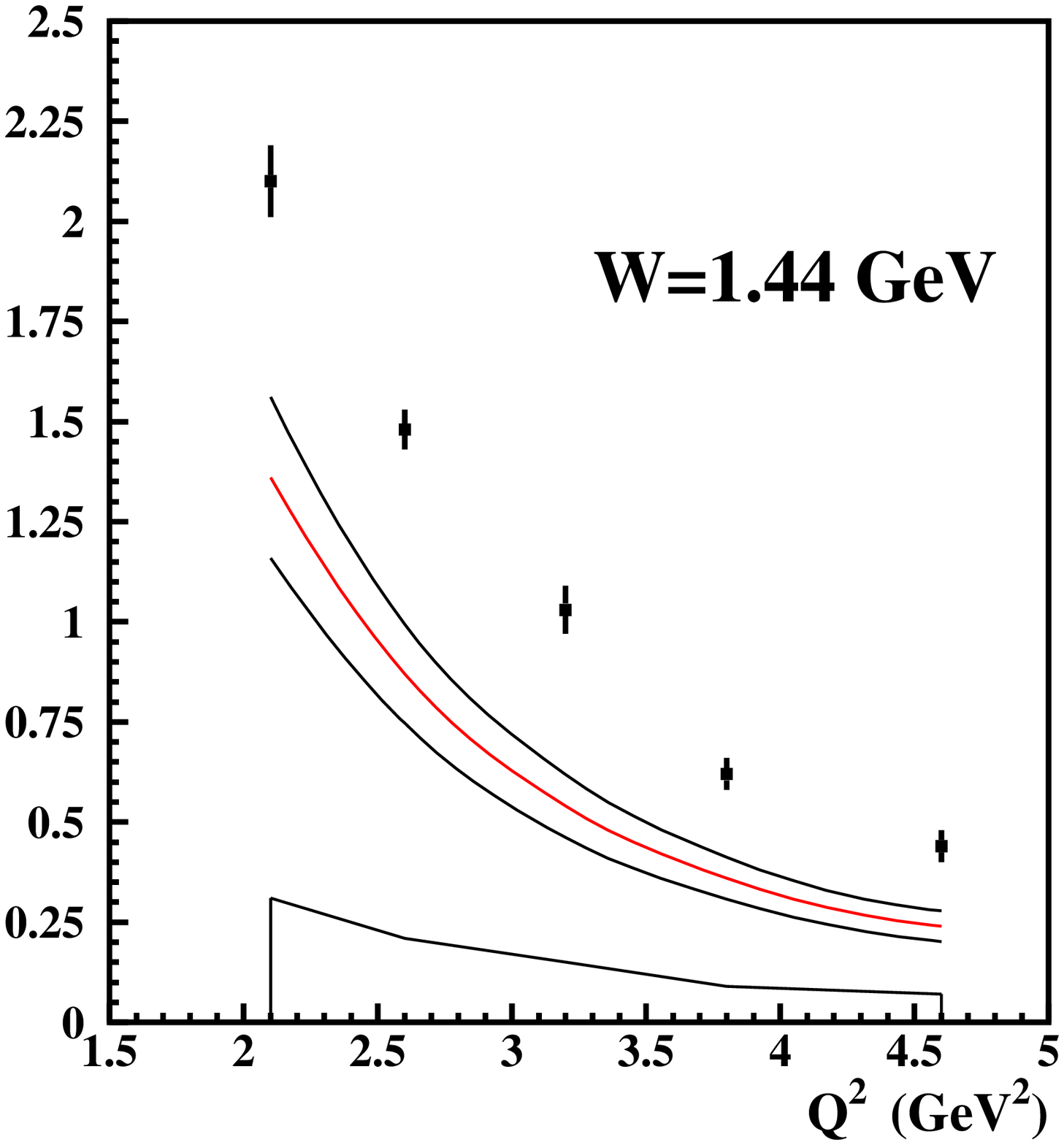}
\includegraphics[width=5.5cm]{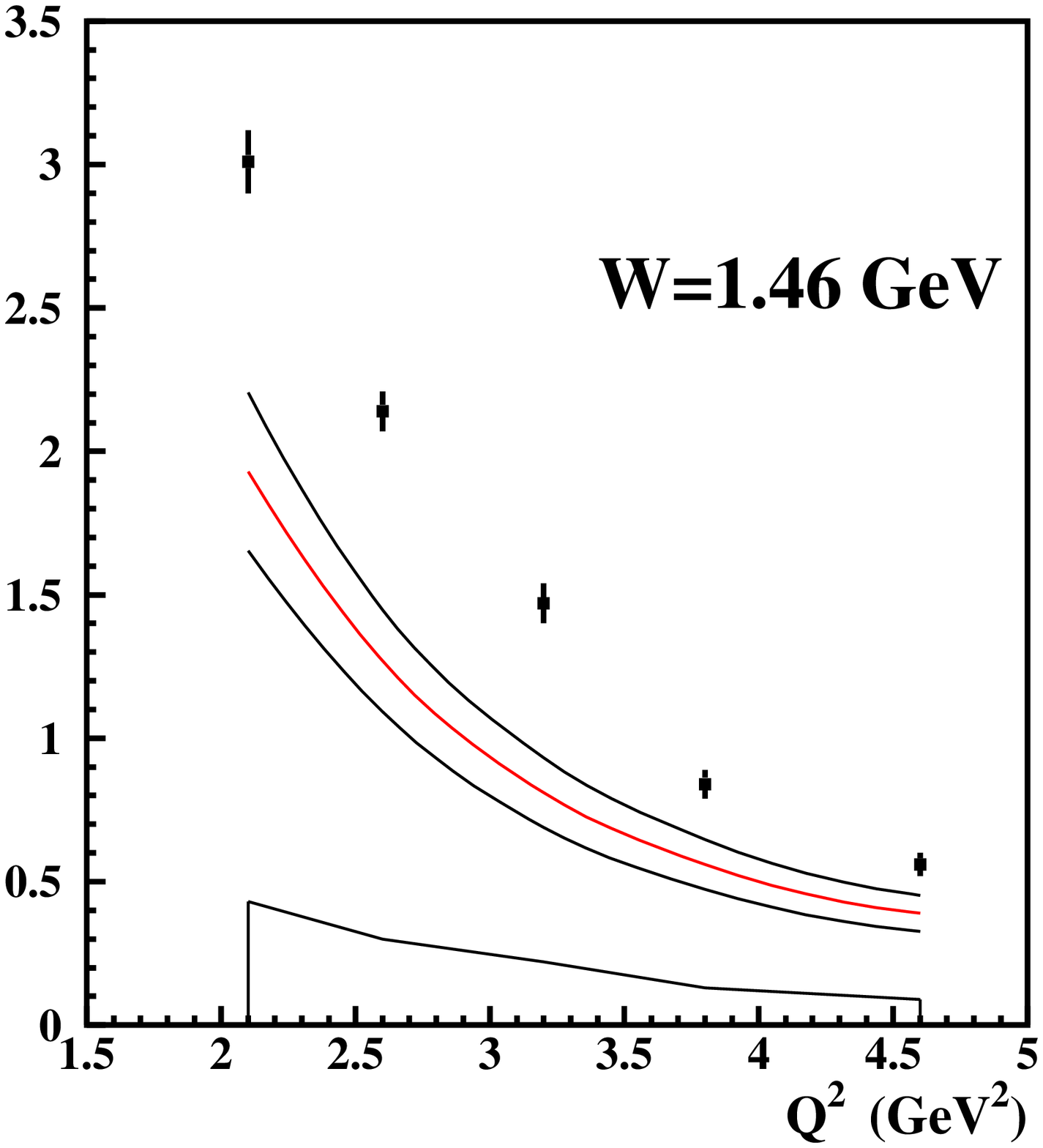}
\vspace{-0.1cm}
\caption{The resonant contributions from the JM16 model~\cite{Mo12,Mo16,Mo16a} computed as 
described in Section~\ref{impact} (red solid lines) in comparison with the CLAS results on 
the fully integrated $\pi^+\pi^-p$ electroproduction cross sections off protons (points with 
statistical error bars) in three $W$ bins near the central mass of the $N(1440)1/2^+$: 
$W$=1.41~GeV (left), $W$=1.44~GeV (center), and $W$=1.46~GeV (right). The systematic
uncertainties of the measurements are shown by the bands at the bottom of each plot. The 
black lines that form a band about the central red JM16 prediction represent the model 
uncertainties.}
\label{xintq214146}
\end{center}
\end{figure*} 

\begin{figure*}[htp]
\begin{center}
\includegraphics[width=5.5cm]{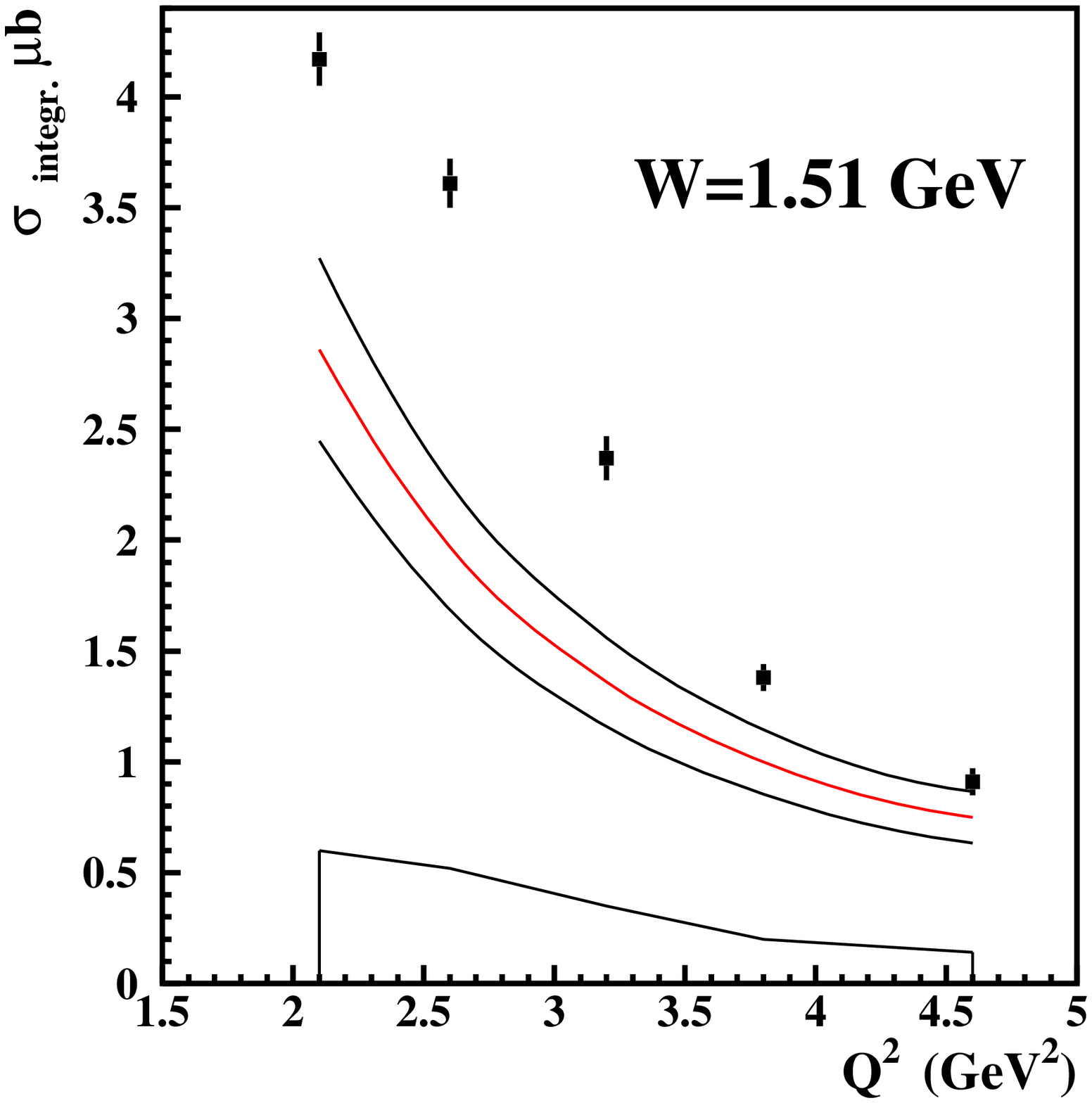}
\includegraphics[width=5.5cm]{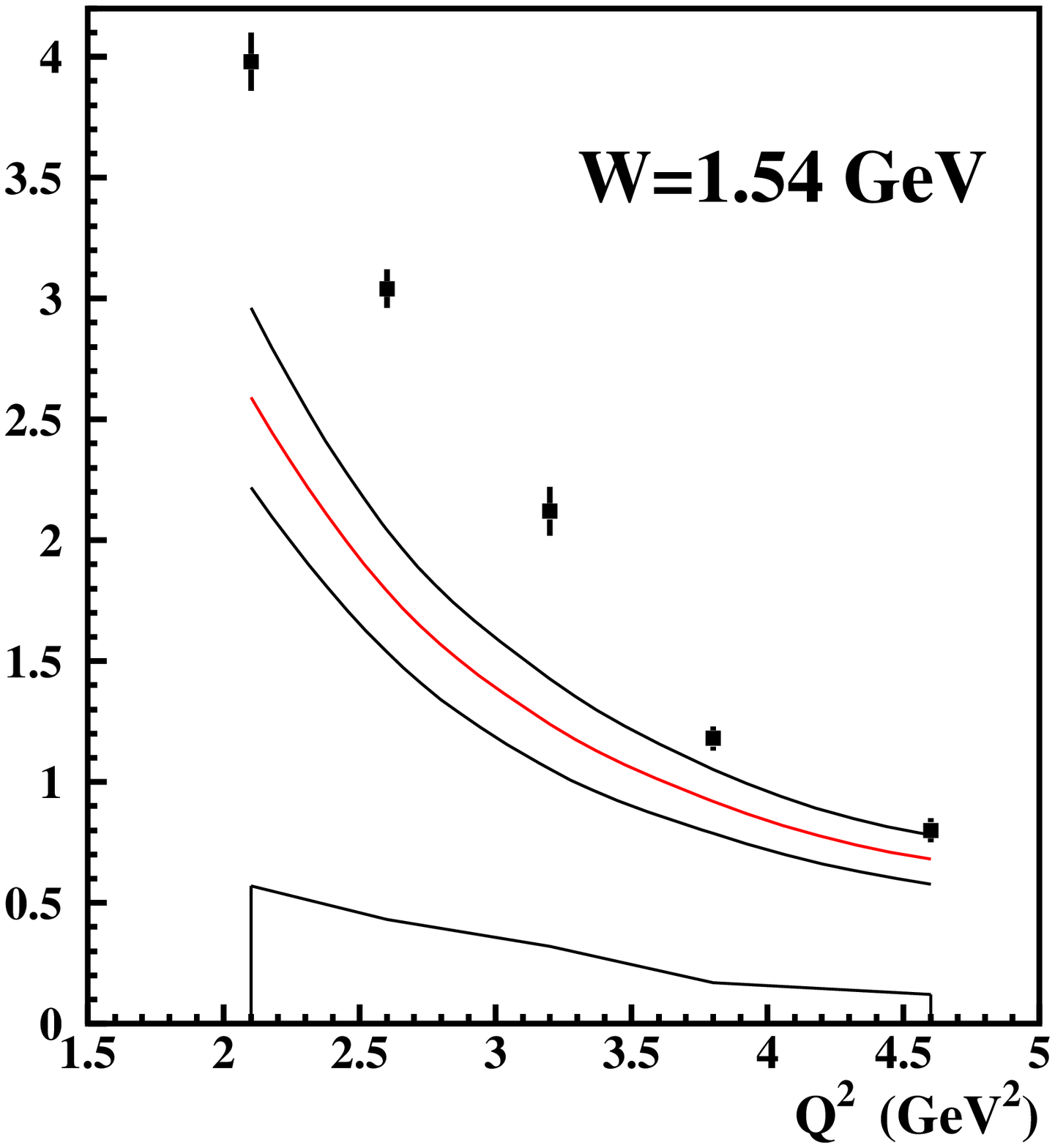}
\includegraphics[width=5.5cm]{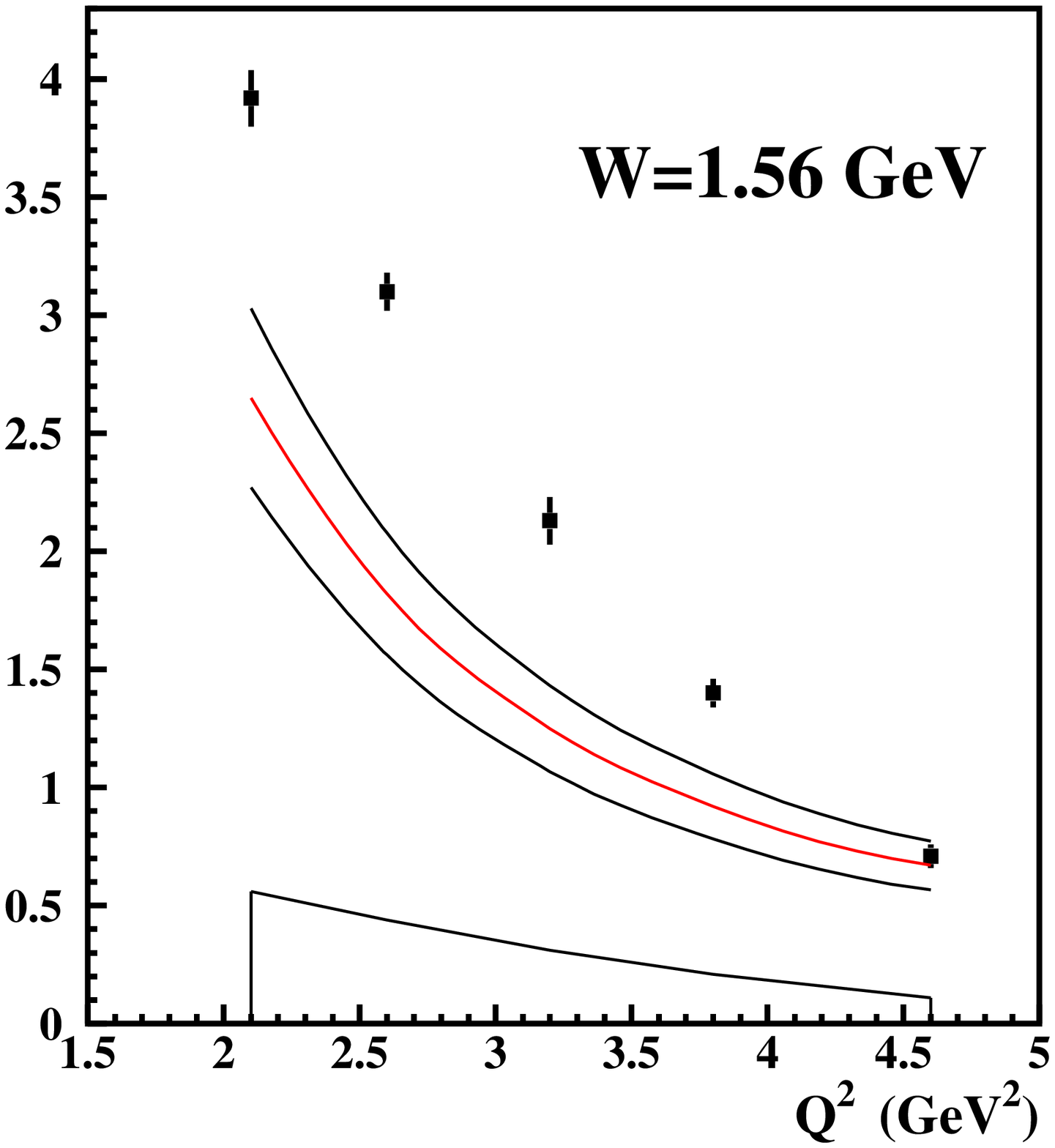}
\vspace{-0.1cm}
\caption{The resonant contributions from the JM16 model~\cite{Mo12,Mo16,Mo16a} computed as 
described in Section~\ref{impact} (red solid lines) in comparison with the CLAS results on 
the fully integrated $\pi^+\pi^-p$ electroproduction cross sections off protons (points with 
statistical error bars) in three $W$ bins near the central mass of the $N(1520)3/2^-$: 
$W$=1.51~GeV (left), $W$=1.54~GeV (center), and $W$=1.56~GeV (right). The systematic
uncertainties of the measurements are shown by the bands at the bottom of each plot. The 
black lines that form a band about the central red JM16 prediction represent the model 
uncertainties.}
\label{xintq2151156}
\end{center}
\end{figure*} 

\begin{figure*}[htp]
\begin{center}
\includegraphics[width=8.91cm]{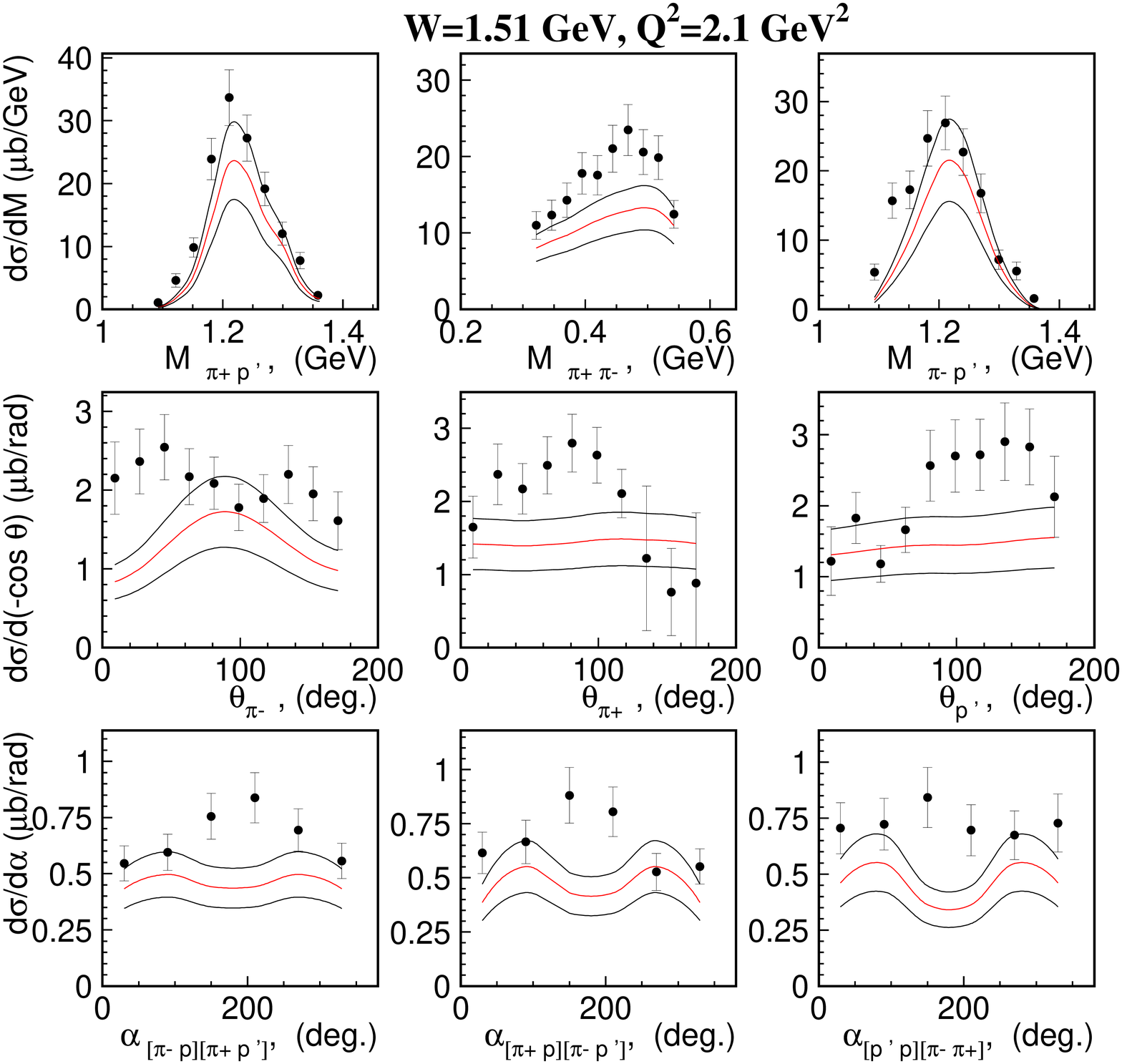}
\includegraphics[width=8.91cm]{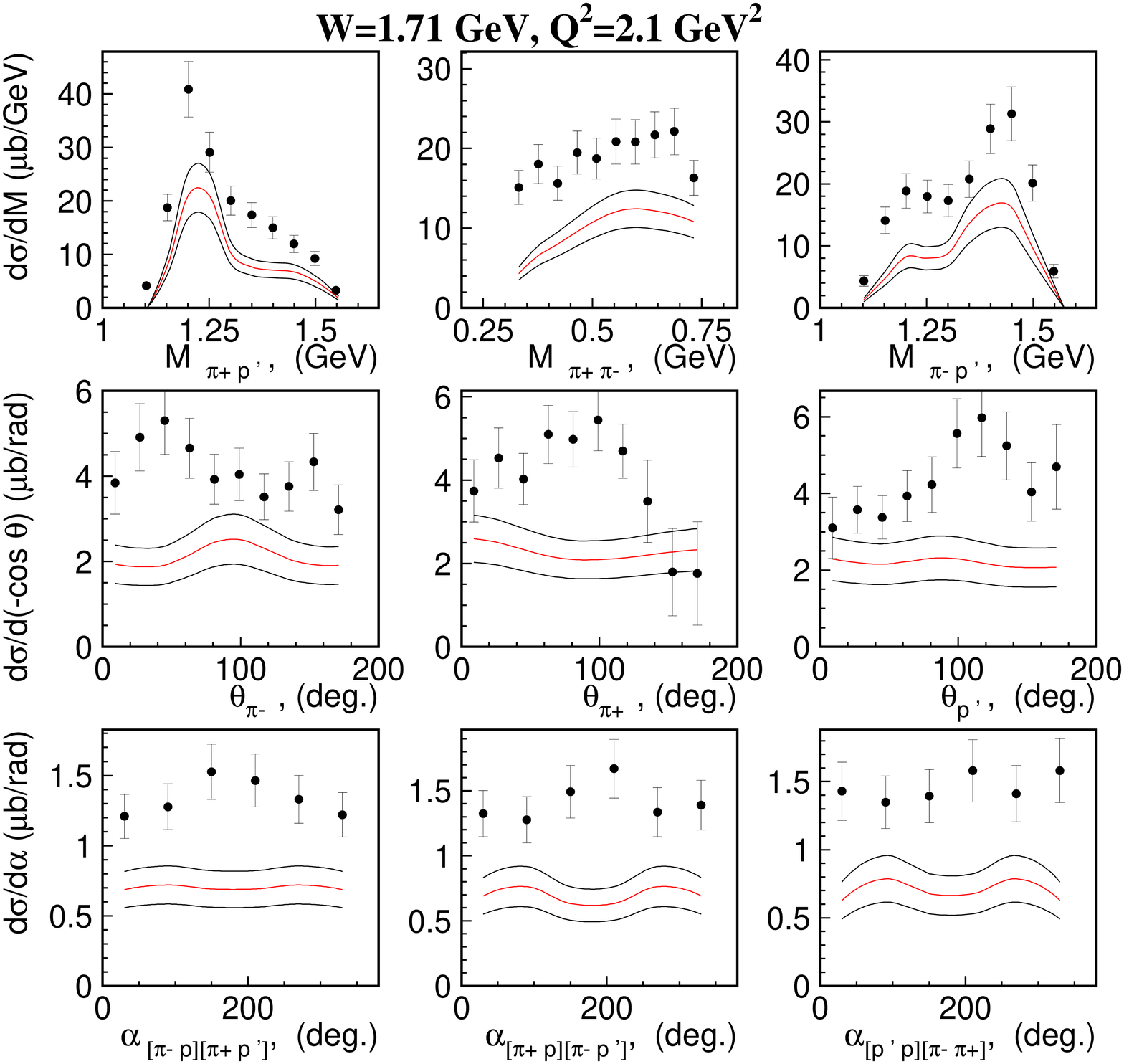}
\vspace{-0.1cm}
\caption{The resonant contributions from the JM16 model~\cite{Mo12,Mo16,Mo16a} (red solid lines) 
to the nine 1-fold differential $\pi^+\pi^-p$ electroproduction cross sections in representative 
$W$ bins inside two $W$ intervals of distinctively different resonant content described in
Section~\ref{impact} at $Q^2$=2.1~GeV$^2$. The black lines that form a band about the central 
red JM16 prediction represent the model uncertainties.}
\label{1diff21}
\end{center}
\end{figure*} 

\begin{figure*}[htp]
\begin{center}
\includegraphics[width=8.91cm]{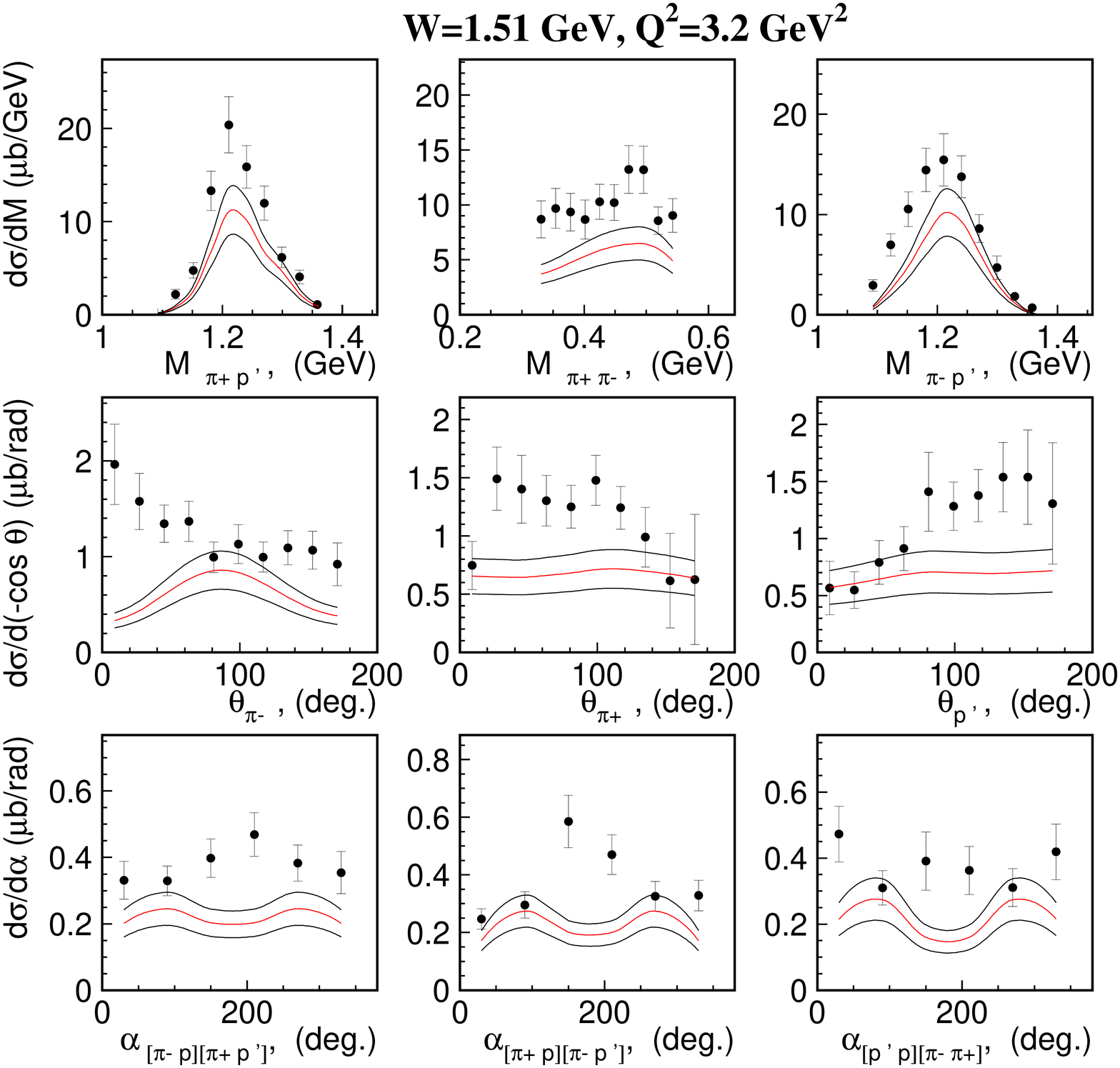}
\includegraphics[width=8.91cm]{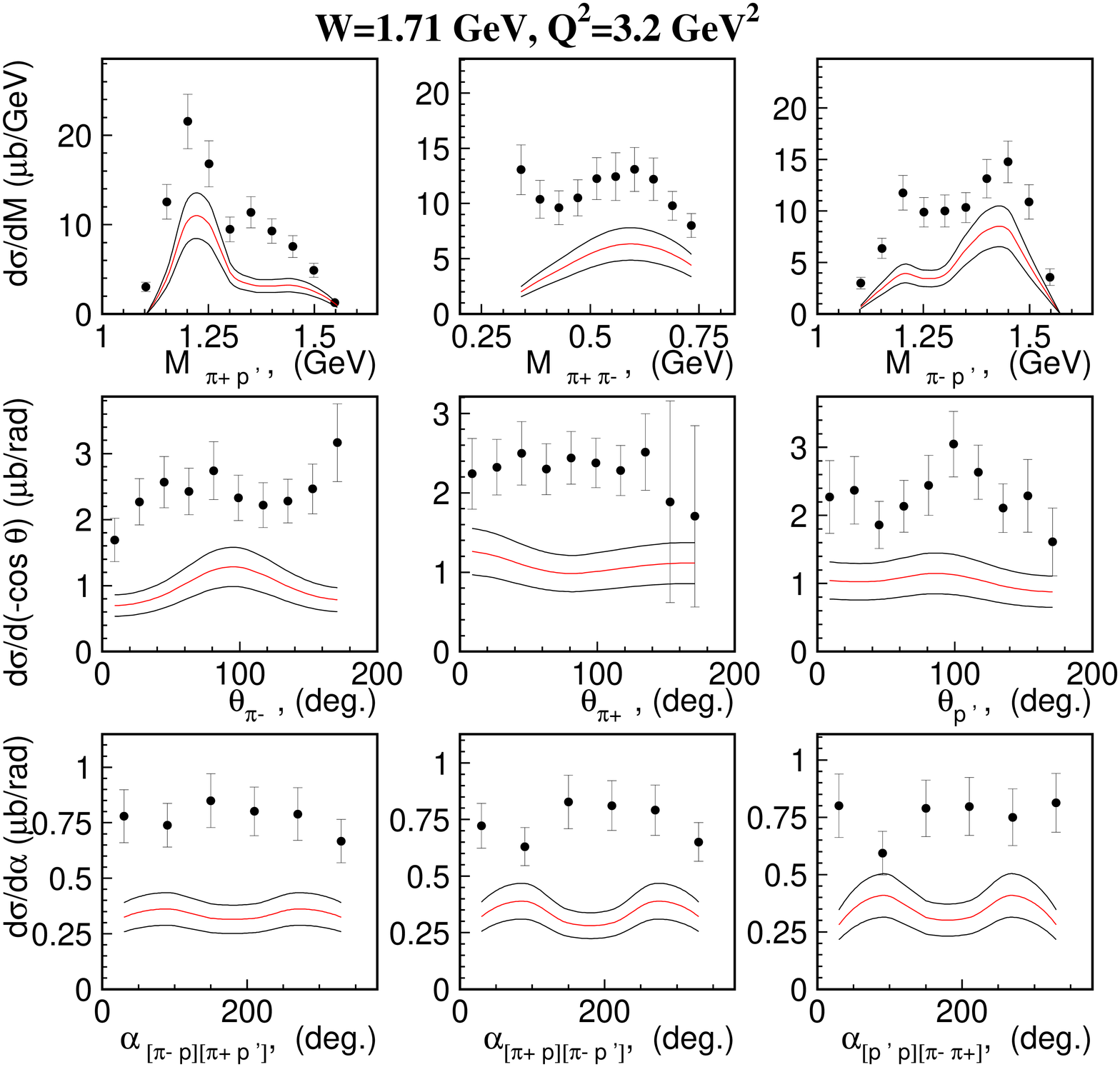}
\vspace{-0.1cm}
\caption{The resonant contributions from the JM16 model~\cite{Mo12,Mo16,Mo16a} (red solid lines) 
to the nine 1-fold differential $\pi^+\pi^-p$ electroproduction cross sections in representative 
$W$ bins inside two $W$ intervals of distinctively different resonant content described in
Section~\ref{impact} at $Q^2$=3.2~GeV$^2$. The black lines that form a band about the central 
red JM16 prediction represent the model uncertainties.}
\label{1diff32}
\end{center}
\end{figure*} 

\begin{figure*}[htp]
\begin{center}
\includegraphics[width=8.91cm]{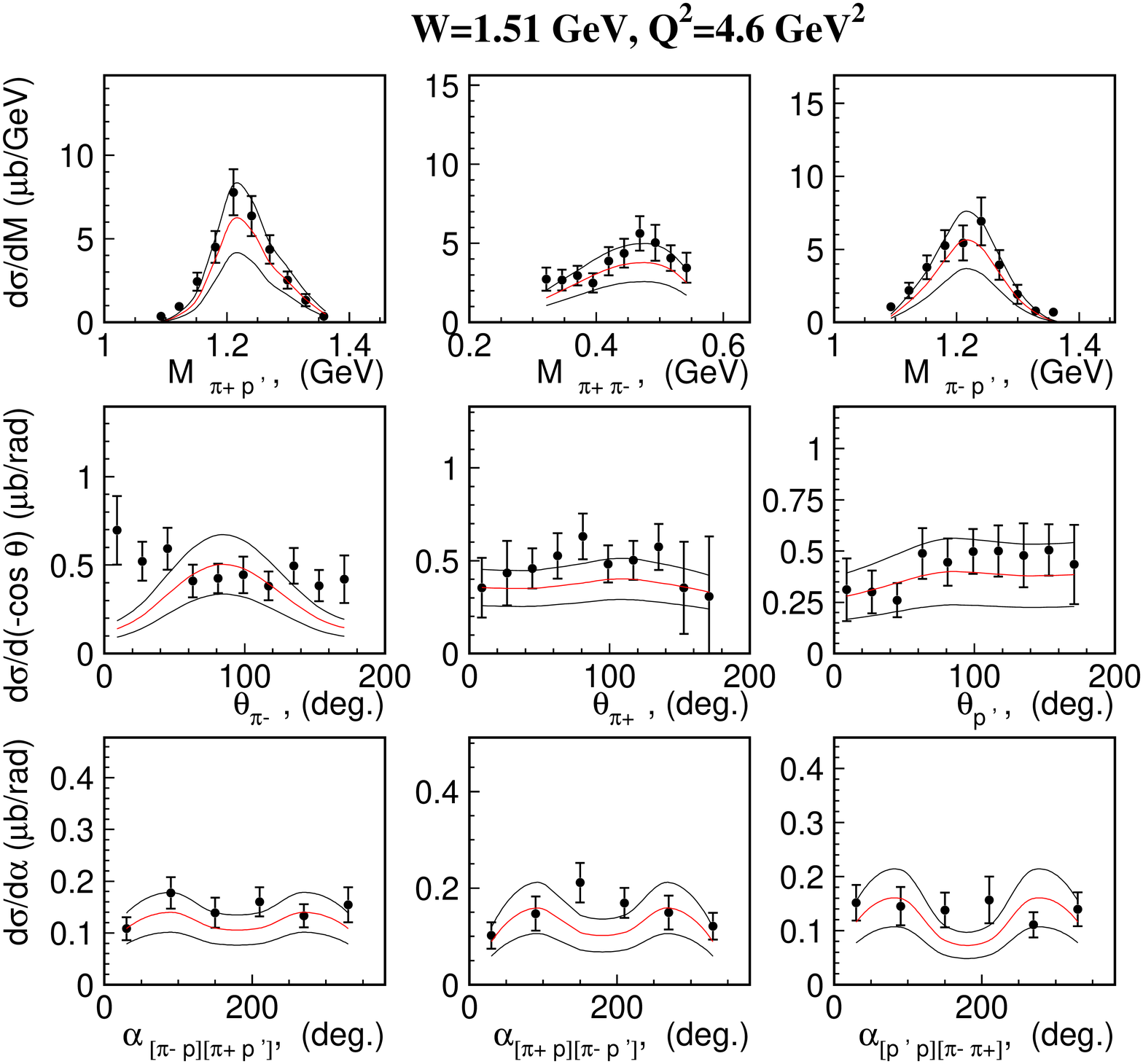}
\includegraphics[width=8.91cm]{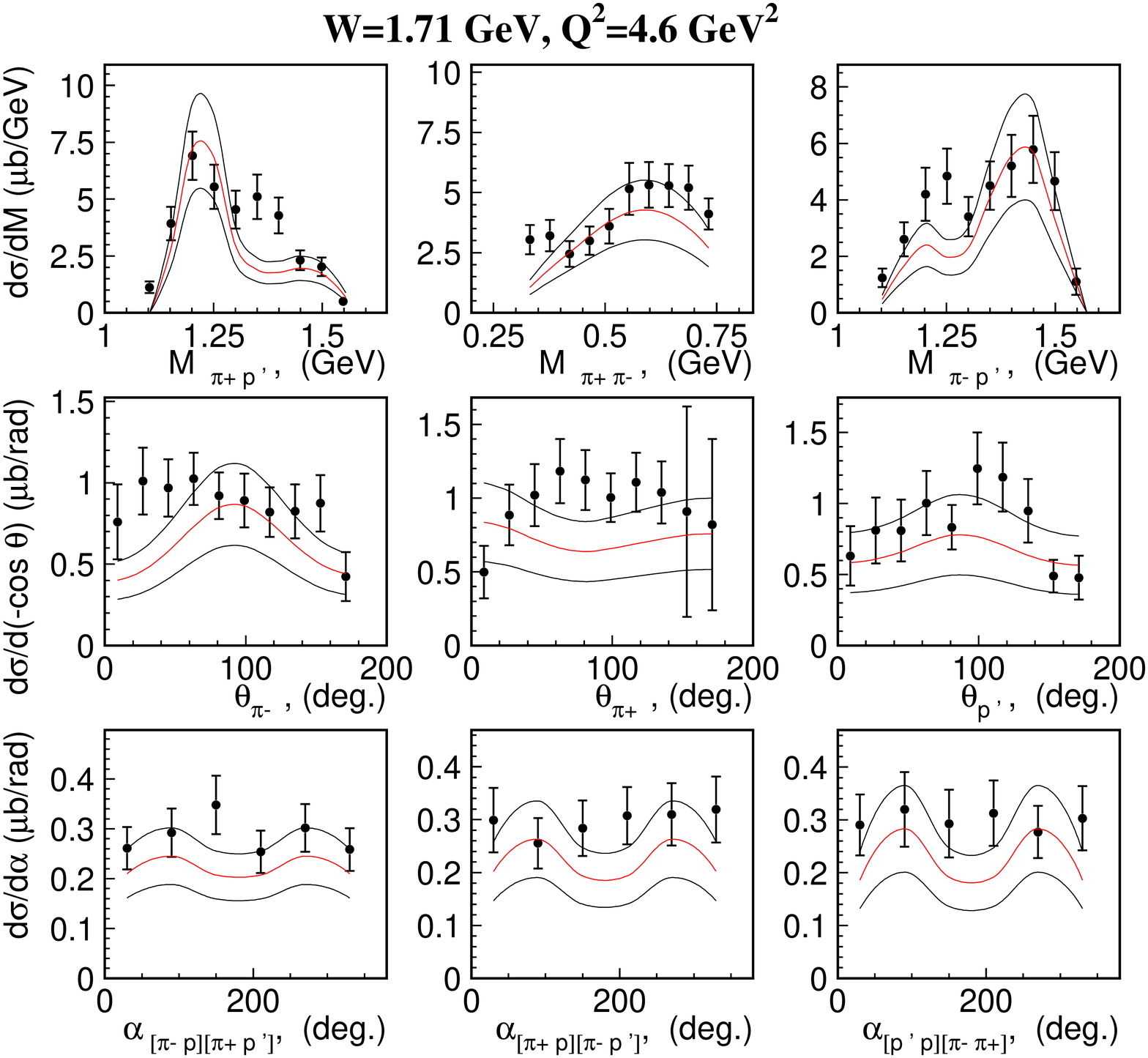}
\vspace{0.5cm}
\caption{The resonant contributions from the JM16 model~\cite{Mo12,Mo16,Mo16a} (red solid lines) 
to the nine 1-fold differential $\pi^+\pi^-p$ electroproduction cross sections in representative 
$W$ bins inside two $W$ intervals of distinctively different resonant content described in
Section~\ref{impact} at $Q^2$=4.6~GeV$^2$. The black lines that form a band about the central 
red JM16 prediction represent the model uncertainties.}
\label{1diff46}
\end{center}
\end{figure*} 

The $Q^2$ dependence of the resonance contributions to the fully integrated $\pi^+\pi^-p$
electroproduction cross sections are shown in Figs.~\ref{xintq214146} and \ref{xintq2151156}. 
The data shown correspond to the $W$ ranges that are closest to the central masses 
of the $N(1440)1/2^+$ and $N(1520)3/2^-$. The electrocouplings of these low-lying 
resonances, as well as for the $N(1535)1/2^-$, are available in the entire range of 
$Q^2$ covered in our measurements~\cite{Den07,Az09,Mo12,Park15,Mo16}. Interpolated values of 
these electrocouplings were used in the resonant contribution evaluations shown in 
Figs.~\ref{xintq214146} and~\ref{xintq2151156}. In the mass range from 1.50~GeV to 1.56~GeV, 
there is also a small contribution from the tail of the $\Delta(1620)1/2^-$ resonance. 
Electrocouplings of this resonance are available at $Q^2 < 1.5$~GeV$^2$~\cite{Mo16}. To 
evaluate this  contribution, the CLAS results were extrapolated into the range 
2.0~GeV$^2$ $< Q^2 <  5.0$~GeV$^2$.

The uncertainties of the resonant contributions were estimated from the quadrature sum of the
statistical and systematic uncertainties of the measured integrated cross sections, assuming 
that the relative uncertainties both for the fully integrated and all 1-fold differential 
cross sections were the same for the measured cross sections and for the computed resonant 
contributions, as was found in previous analyses of $\pi^+\pi^-p$ electroproduction data from 
CLAS~\cite{Mo12,Mo16}. Under this assumption, the initial evaluation of the uncertainties for 
the resonant contributions was performed accounting for only statistical uncertainties of the 
measured integrated and 1-fold differential cross sections. However, the statistical 
uncertainties offer a reasonable estimate only in the case when the $\chi^2/d.p.$ 
($\chi^2$ per data point) achieved in the data fit is close to unity. The $\chi^2/d.p.$ values 
achieved in the previous analyses of the CLAS $\pi^+\pi^-p$ electroproduction data were in the 
range from 1.3 to 2.9~\cite{Mo12,Mo16,Mo16a}. In order to account for the additional data 
uncertainties responsible for the deviation of the $\chi^2/d.p.$ values from unity, we 
multiplied the initial values of the uncertainties for the resonant contributions by the root 
square of the averaged $\chi^2/d.p.$ value achieved in the previous data fits, which was equal 
to 1.45.  Uncertainties of the estimated resonant contributions to the fully integrated 
$\pi^+\pi^-p$ electroproduction cross sections are represented in Figs.~\ref{xintq214146} and 
\ref{xintq2151156} by the areas between the black solid lines.
 
The results shown in  Figs.~\ref{xintq214146} and \ref{xintq2151156} demonstrate an increase 
with $Q^2$ of the relative resonance contributions to the fully integrated $\pi^+\pi^-p$ 
electroproduction cross sections. The resonant part begins to dominate at $Q^2 > 4.0$~GeV$^2$. 
Table~\ref{ratio} shows ratios of the projected resonant contributions to the measured cross 
sections in several $Q^2$ bins averaged within three $W$ intervals that have distinctively 
different resonant content.
\begin{itemize}
\item In the interval 1.41~GeV $< W < 1.61$~GeV, electrocouplings of the low-lying resonances 
have been measured in the $Q^2$ range covered here.

\item For the states in the mass range 1.61~GeV $<  W < 1.74$~GeV that contribute to the 
$\pi^+\pi^-p$ electroproduction, only electrocouplings of the $N(1685)5/2^+$ resonance are 
available from the CLAS $\pi N$ data~\cite{Park15} in the range of $Q^2$ covered in our 
measurements. The $\Delta(1620)1/2^-$, $\Delta(1700)3/2^-$, $N(1720)3/2^+$, and candidate
$N'(1720)3/2^+$ states decay preferentially to $\pi\pi N$. Their contributions, as well as 
from the $N(1650)1/2^-$ to the $\pi^+\pi^-p$ cross sections, have been evaluated by 
extrapolating the available electrocouplings from $Q^2 <1.5$~GeV$^2$~\cite{Mo16a} to 
2.0~GeV$^2 < Q^2 < 5.0$~GeV$^2$.

\item The interval 1.74~GeV $< W < 1.82$~GeV includes only states recently reported~\cite{Rpp14} 
for which no electrocouplings are available to date, and their $\pi\pi N$ couplings are also 
unknown. Hence no projections are possible in this mass range. No resonances in this mass range 
were included for evaluation of the resonant contributions to the $\pi^+\pi^-p$ electroproduction 
cross sections. 
\end{itemize}


\begin{table*}
\begin{center}
\begin{tabular}{|c|c|c|c|c|} \hline
$Q^2$,      & 1.41 $<$ $W$ $<$ 1.61, & 1.61 $<$ $W$ $<$ 1.74, &  1.74 $<$ $W$ $<$ 1.82, \\
GeV$^2$     &~GeV                  &~GeV                      & ~GeV                    \\ \hline
2.1         & 0.650 $\pm$ 0.033     & 0.570 $\pm$ 0.034         & 0.200 $\pm$ 0.019     \\
2.6         & 0.570 $\pm$ 0.029     & 0.500 $\pm$ 0.028         & 0.180 $\pm$ 0.010     \\
3.2         & 0.550 $\pm$ 0.029     & 0.490 $\pm$ 0.029         & 0.190 $\pm$ 0.017     \\
3.8         & 0.660 $\pm$ 0.034     & 0.620 $\pm$ 0.034         & 0.210 $\pm$ 0.014     \\
4.6         & 0.750 $\pm$ 0.041     & 0.790 $\pm$ 0.049         & 0.240 $\pm$ 0.017     \\ \hline
\end{tabular}
\caption{Ratios of the resonant contributions computed within the framework of the current 
JM16 model version~\cite{Mo12,Mo16,Mo16a} relative to the measured fully integrated 
$e p \to e' \pi^+\pi^-p'$ cross sections averaged within three $W$ intervals with different 
resonant content.}
\label{ratio}  
\end{center}
\end{table*}

In Figs.~\ref{1diff21},~\ref{1diff32}, and \ref{1diff46} we show the comparison of the 
nine 1-fold differential $\pi^+\pi^-p$ electroproduction cross sections and the resonant 
contributions computed in the JM16 model within the given $W$ and $Q^2$ bins. The resonant 
contributions obtained with the resonant parameters of the JM16 model taken from previous 
analyses of the CLAS $\pi^+\pi^-p$ electroproduction data at $Q^2 < 1.5$~GeV$^2$
\cite{Mo12,Mo16} after interpolation/extrapolation of the $\gamma_vpN^*$ electrocouplings 
to the $Q^2$ range covered in our measurements, are shown by the red lines. The uncertainties 
for the resonant contributions were evaluated as described above. The procedure 
for the evaluation of the resonant contributions to the 1-fold differential cross sections 
within the framework of the unitarized Breit-Wigner ansatz is described in \cite{Mo12,Mo16}. 
The uncertainties in the resonant contributions to the 1-fold differential cross sections  
are shown in Figs.~\ref{1diff21}, \ref{1diff32}, and \ref{1diff46} by the areas 
between the black solid lines.  

According to the results in Figs.~\ref{1diff21}, \ref{1diff32}, and \ref{1diff46}, the 
projected resonance contributions to the measured cross sections at $W < 1.74$~GeV are the 
largest over the entire $Q^2$ range covered here as shown in Table~\ref{ratio}. We find that the 
relative resonant contributions increase with $Q^2$ and dominate the integrated cross section 
in the highest $Q^2$ bin centered at 4.6~GeV$^2$. 

However, the resonant contributions to the CM angular distributions at $Q^2=4.6$~GeV$^2$ and 
in the mass range 1.51~GeV to 1.71~GeV shown in Fig.~\ref{1diff46} indicate sizable differences 
in the angular dependence of the measured differential cross sections and the projected 
resonance contributions. This suggests substantial contributions from non-resonant mechanisms 
even at the highest photon virtualities covered by our measurements.
 
In particular, a comparison of the measured CM angular distributions for the final state 
$\pi^-$ and the computed resonant contributions shown in Fig.~\ref{1diff46} suggests that 
the non-resonant contribution from the $\pi^-\Delta^{++}$ intermediate state created in the 
$t$-channel exchange dominates at forward angles. Also, the presence of a direct $2\pi$ 
production mechanism may explain the differences between the measured cross sections and 
the resonant contributions seen at the backward $\pi^-$ angles.  

In the $W$ interval from 1.74~GeV to 1.82~GeV the ratio of the projected resonant 
contributions to the fully integrated $\pi^+\pi^- p$ electroproduction cross sections 
decreases by more than a factor of two in all $Q^2$ bins covered here (Table~\ref{ratio}). 
In order to achieve a satisfactory description of the data in this mass range with the
resonant contributions from the aforementioned resonances only, requires an increase of the 
relative contribution from the non-resonant mechanisms by more than a factor of two, which 
seems unlikely. 


The data discussed here therefore present an opportunity to independently verify signals 
from new baryon states reported in the Bonn-Gatchina photoproduction data analysis 
\cite{Sar12}. A successful description of the $\pi^+\pi^-p$ photo- and electroproduction 
data with $Q^2$-independent resonance parameters (such as partial $\pi \Delta$ and $\rho p$ 
decay widths) would provide strong evidence for these newly claimed excited 
nucleon states. 

According to Table~\ref{ratio}, at $W$ $<$ 1.74~GeV the relative resonant contributions 
decrease in the $Q^2$ range from 2.0~GeV$^2$ to 3.0~GeV$^2$, while at $Q^2 > 3.0$~GeV$^2$ 
the relative resonant contributions exhibit an increase with $Q^2$. For resonances in the 
mass range from 1.41~GeV to 1.61~GeV, the electrocouplings are known from CLAS data in the 
entire range of photon virtualities covered by our measurements. Therefore, this effect 
cannot be related to uncertainties resulting from the extrapolations of the resonance 
electrocouplings.

Our data suggest that at $Q^2 < 3.0$~GeV$^2$ the resonance contributions decrease with 
$Q^2$ faster in comparison with other contributing mechanisms. Instead, at $Q^2 > 3.0$~GeV$^2$ 
the resonance contributions decrease with $Q^2$ slower in comparison with the remaining 
contributions to exclusive $\pi^+\pi^-p$ electroproduction. Such behavior supports the 
assessment of the structure of the $N^*$ states from analyses of exclusive meson 
electroproduction~\cite{Bu12,Mo16} as an interplay of the inner core of three dressed quarks 
and the external meson-baryon cloud. The range of $Q^2 < 3.0$~GeV$^2$ corresponds to 
substantial contributions from the meson-baryon cloud, which becomes largest at the photon 
point. This contribution decreases with $Q^2$ faster than the contribution from non-resonant 
mechanisms and its relative resonant contribution decreases with $Q^2$ for $Q^2 < 3.0$~GeV$^2$. 
Instead, at higher $Q^2$ the contribution from the quark core becomes more significant, even 
dominant, and this contribution decreases with $Q^2$ more slowly than the non-resonant processes, 
causing relative growth of the resonant cross sections. 

\section{CONCLUSIONS}

In this paper we presented new electroproduction data on $ep \to e \pi^+\pi^-p'$ in the mass 
range $W < 2.0$~GeV, and at photon virtualities 2.0~GeV$^2$ $< Q^2 < 5.0$~GeV$^2$. The 
kinematics covered is rich with known nucleon resonances whose electrocouplings are either 
unknown or known from $\pi N$ electroproduction only. In particular, these data cover the 
range of $W > 1.6$~GeV, where many resonances  couple predominantly to the $\pi\pi N$ final 
state, and hence can be studied here. 

The extraction of the electrocoupling amplitudes requires a reaction model that must include 
all well established resonances in amplitude form, along with the amplitudes of the relevant 
non-resonant mechanisms and the interference of the contributing amplitudes. One such model 
is the JM framework~\cite{Mo12,Mo16,Mo16a}, but its reach in the invariant mass of the final 
hadrons $W$ and photon virtuality $Q^2$ must be extended into the kinematic domain of the 
new data. This effort is underway and the results will be part of a future publication on the 
subject. 

The projected resonant contributions to the cross sections discussed in Section~\ref{impact} 
were obtained within the framework of the unitarized Breit-Wigner ansatz of the JM16 
version of the JM model~\cite{Mo12}. The resonant cross sections were evaluated with 
electrocouplings determined by interpolations and extrapolations of the available results on 
these resonance parameters~\cite{Is16,resnum} from the measured $Q^2$ into new territory. 

Our studies show strong indications that the relative contributions of the resonant cross 
section at $W < 1.74$~GeV increase with $Q^2$. This suggests good prospects for the exploration 
of electrocouplings of the nucleon resonances in this mass range and with photon 
virtualities up to 5.0~GeV$^2$ and above. With the CEBAF accelerator upgrade to an energy of 
12~GeV and by employing the new CLAS12 detector, photon virtualities in the range 
5.0~GeV$^2 < Q^2 < 12.0$~GeV$^2$ can be reached for all of the prominent resonances with 
masses below 2.0~GeV. The range of $Q^2 > 2.0$~GeV$^2$ is of particular importance to study 
the momentum dependence of the light-quark masses, as the $Q^2$ dependence of the resonance 
electrocouplings has been shown to be sensitive to the quark mass function~\cite{Cr16,Cr16a}. 
This provides a sensitive means of testing computations of the electrocouplings from first 
principles QCD as incorporated in the Dyson-Schwinger equation (DSE) approach~\cite{Seg14,Seg15}.

The data presented here provide a basis to verify the existence of possible new baryon 
states reported at $M > 1.8$~GeV in a global multi-channel partial wave analysis of
photoproduction data by the 
Bonn-Gatchina group~\cite{Sar14}. The apparent decrease in the resonant contributions 
at $W > 1.74$~GeV, as shown in Table~\ref{ratio}, suggests that more 
resonances in this mass range will be needed to describe the present data, as well as 
the possibility to locate new baryon states by examining these data 
with $Q^2$ independent hadronic parameters for the excited nucleon states.  
In addition, reaching higher mass states at 2~GeV and above will allow us to test the quark 
model predictions employing light-front dynamics~\cite{Az15} and other approaches~\cite{San15} 
in a domain where first principles calculations are still unavailable. 

\begin{acknowledgments}
We are grateful for theoretical motivation and support of our experiment by I.G. Aznauryan, V.M. 
Braun, C.D. Roberts, E. Santopinto. 
We express our gratitude for the efforts of the staff of the Accelerator and Physics 
Divisions at Jefferson Lab that made this experiment possible. This work was supported 
in part by the U.S. Department of Energy (DOE) and National Science Foundation (NSF),
the Chilean Comisi\'on Nacional de Investigaci\'on Cient\'ifica y Tecnol\'ogica (CONICYT),
the Italian Istituto Nazionale di Fisica Nucleare (INFN),
the French Centre National de la Recherche Scientifique (CNRS),
the French Commissariat \`{a} l'Energie Atomique (CEA), 
the Skobeltsyn Institute of Nuclear Physics (SINP) and the Physics Departments at Moscow State 
University (MSU, Moscow) and Ohio University (OU), the Scottish Universities Physics Alliance 
(SUPA), the National Research Foundation of Korea (NRF), 
the UK Science and Technology Facilities Council (STFC). 
Jefferson Science Associates (JSA) operates the Thomas Jefferson National Accelerator 
Facility for the United States Department of Energy under contract DE-AC05-06OR23177. 
\end{acknowledgments}

\bibliography{2pi_vm}{}

\begin{thebibliography}{38}%
\makeatletter
\providecommand \@ifxundefined [1]{%
 \@ifx{#1\undefined}
}%
\providecommand \@ifnum [1]{%
 \ifnum #1\expandafter \@firstoftwo
 \else \expandafter \@secondoftwo
 \fi
}%
\providecommand \@ifx [1]{%
 \ifx #1\expandafter \@firstoftwo
 \else \expandafter \@secondoftwo
 \fi
}%
\providecommand \natexlab [1]{#1}%
\providecommand \enquote  [1]{``#1''}%
\providecommand \bibnamefont  [1]{#1}%
\providecommand \bibfnamefont [1]{#1}%
\providecommand \citenamefont [1]{#1}%
\providecommand \href@noop [0]{\@secondoftwo}%
\providecommand \href [0]{\begingroup \@sanitize@url \@href}%
\providecommand \@href[1]{\@@startlink{#1}\@@href}%
\providecommand \@@href[1]{\endgroup#1\@@endlink}%
\providecommand \@sanitize@url [0]{\catcode `\\12\catcode `\$12\catcode
  `\&12\catcode `\#12\catcode `\^12\catcode `\_12\catcode `\%12\relax}%
\providecommand \@@startlink[1]{}%
\providecommand \@@endlink[0]{}%
\providecommand \url  [0]{\begingroup\@sanitize@url \@url }%
\providecommand \@url [1]{\endgroup\@href {#1}{\urlprefix }}%
\providecommand \urlprefix  [0]{URL }%
\providecommand \Eprint [0]{\href }%
\providecommand \doibase [0]{http://dx.doi.org/}%
\providecommand \selectlanguage [0]{\@gobble}%
\providecommand \bibinfo  [0]{\@secondoftwo}%
\providecommand \bibfield  [0]{\@secondoftwo}%
\providecommand \translation [1]{[#1]}%
\providecommand \BibitemOpen [0]{}%
\providecommand \bibitemStop [0]{}%
\providecommand \bibitemNoStop [0]{.\EOS\space}%
\providecommand \EOS [0]{\spacefactor3000\relax}%
\providecommand \BibitemShut  [1]{\csname bibitem#1\endcsname}%
\let\auto@bib@innerbib\@empty
\bibitem [{\citenamefont {Aznauryan}\ and\ \citenamefont
  {Burkert}(2012)}]{Bu12}%
  \BibitemOpen
  \bibfield  {author} {\bibinfo {author} {\bibfnamefont {I.~G.}\ \bibnamefont
  {Aznauryan}}\ and\ \bibinfo {author} {\bibfnamefont {V.~D.}\ \bibnamefont
  {Burkert}},\ }\href {\doibase http://dx.doi.org/10.1016/j.ppnp.2011.08.001}
  {\bibfield  {journal} {\bibinfo  {journal} {Prog. Part. Nucl. Phys.}\
  }\textbf {\bibinfo {volume} {67}},\ \bibinfo {pages} {1 } (\bibinfo {year}
  {2012})}\BibitemShut {NoStop}%
\bibitem [{\citenamefont {Aznauryan}\ \emph {et~al.}(2013)\citenamefont
  {Aznauryan} \emph {et~al.}}]{Az13}%
  \BibitemOpen
  \bibfield  {author} {\bibinfo {author} {\bibfnamefont {I.~G.}\ \bibnamefont
  {Aznauryan}} \emph {et~al.},\ }\href {\doibase 10.1142/S0218301313300154}
  {\bibfield  {journal} {\bibinfo  {journal} {Int. J. Mod. Phys. E}\ }\textbf
  {\bibinfo {volume} {22}},\ \bibinfo {pages} {1330015} (\bibinfo {year}
  {2013})}\BibitemShut {NoStop}%
\bibitem [{\citenamefont {{Aznauryan}}\ \emph {et~al.}()\citenamefont
  {{Aznauryan}}, \citenamefont {{Burkert}},\ and\ \citenamefont
  {{Mokeev}}}]{Bu15}%
  \BibitemOpen
  \bibfield  {author} {\bibinfo {author} {\bibfnamefont {I.~G.}\ \bibnamefont
  {{Aznauryan}}}, \bibinfo {author} {\bibfnamefont {V.~D.}\ \bibnamefont
  {{Burkert}}}, \ and\ \bibinfo {author} {\bibfnamefont {V.~I.}\ \bibnamefont
  {{Mokeev}}},\ }\href@noop {} {\ }\Eprint {http://arxiv.org/abs/1509.08523}
  {arXiv:1509.08523 [nucl-ex]} \BibitemShut {NoStop}%
\bibitem [{\citenamefont {Mokeev}\ \emph
  {et~al.}(2016{\natexlab{a}})\citenamefont {Mokeev} \emph {et~al.}}]{Mo16}%
  \BibitemOpen
  \bibfield  {author} {\bibinfo {author} {\bibfnamefont {V.~I.}\ \bibnamefont
  {Mokeev}} \emph {et~al.},\ }\href {\doibase 10.1103/PhysRevC.93.025206}
  {\bibfield  {journal} {\bibinfo  {journal} {Phys. Rev. C}\ }\textbf {\bibinfo
  {volume} {93}},\ \bibinfo {pages} {025206} (\bibinfo {year}
  {2016}{\natexlab{a}})}\BibitemShut {NoStop}%
\bibitem [{\citenamefont {Aznauryan}\ and\ \citenamefont
  {Burkert}(2015)}]{Az15}%
  \BibitemOpen
  \bibfield  {author} {\bibinfo {author} {\bibfnamefont {I.~G.}\ \bibnamefont
  {Aznauryan}}\ and\ \bibinfo {author} {\bibfnamefont {V.~D.}\ \bibnamefont
  {Burkert}},\ }\href {\doibase 10.1103/PhysRevC.92.015203} {\bibfield
  {journal} {\bibinfo  {journal} {Phys. Rev. C}\ }\textbf {\bibinfo {volume}
  {92}},\ \bibinfo {pages} {015203} (\bibinfo {year} {2015})}\BibitemShut
  {NoStop}%
\bibitem [{\citenamefont {Suzuki}\ \emph {et~al.}(2010)\citenamefont {Suzuki},
  \citenamefont {Sato},\ and\ \citenamefont {Lee}}]{Lee10}%
  \BibitemOpen
  \bibfield  {author} {\bibinfo {author} {\bibfnamefont {N.}~\bibnamefont
  {Suzuki}}, \bibinfo {author} {\bibfnamefont {T.}~\bibnamefont {Sato}}, \ and\
  \bibinfo {author} {\bibfnamefont {T.-S.~H.}\ \bibnamefont {Lee}},\ }\href
  {\doibase 10.1103/PhysRevC.82.045206} {\bibfield  {journal} {\bibinfo
  {journal} {Phys. Rev. C}\ }\textbf {\bibinfo {volume} {82}},\ \bibinfo
  {pages} {045206} (\bibinfo {year} {2010})}\BibitemShut {NoStop}%
\bibitem [{\citenamefont {Braun}\ \emph {et~al.}(2014)\citenamefont {Braun}
  \emph {et~al.}}]{Br14}%
  \BibitemOpen
  \bibfield  {author} {\bibinfo {author} {\bibfnamefont {V.~M.}\ \bibnamefont
  {Braun}} \emph {et~al.},\ }\href {\doibase 10.1103/PhysRevD.89.094511}
  {\bibfield  {journal} {\bibinfo  {journal} {Phys. Rev. D}\ }\textbf {\bibinfo
  {volume} {89}},\ \bibinfo {pages} {094511} (\bibinfo {year}
  {2014})}\BibitemShut {NoStop}%
\bibitem [{\citenamefont {Aznauryan}\ \emph {et~al.}(2009)\citenamefont
  {Aznauryan} \emph {et~al.}}]{Az09}%
  \BibitemOpen
  \bibfield  {author} {\bibinfo {author} {\bibfnamefont {I.~G.}\ \bibnamefont
  {Aznauryan}} \emph {et~al.} (\bibinfo {collaboration} {CLAS Collaboration}),\
  }\href {\doibase 10.1103/PhysRevC.80.055203} {\bibfield  {journal} {\bibinfo
  {journal} {Phys. Rev. C}\ }\textbf {\bibinfo {volume} {80}},\ \bibinfo
  {pages} {055203} (\bibinfo {year} {2009})}\BibitemShut {NoStop}%
\bibitem [{\citenamefont {Anikin}\ \emph {et~al.}(2015)\citenamefont {Anikin},
  \citenamefont {Braun},\ and\ \citenamefont {Offen}}]{Br15}%
  \BibitemOpen
  \bibfield  {author} {\bibinfo {author} {\bibfnamefont {I.~V.}\ \bibnamefont
  {Anikin}}, \bibinfo {author} {\bibfnamefont {V.~M.}\ \bibnamefont {Braun}}, \
  and\ \bibinfo {author} {\bibfnamefont {N.}~\bibnamefont {Offen}},\ }\href
  {\doibase 10.1103/PhysRevD.92.014018} {\bibfield  {journal} {\bibinfo
  {journal} {Phys. Rev. D}\ }\textbf {\bibinfo {volume} {92}},\ \bibinfo
  {pages} {014018} (\bibinfo {year} {2015})}\BibitemShut {NoStop}%
\bibitem [{\citenamefont {Segovia}\ and\ \citenamefont
  {others}(2014)\citenamefont {Segovia} \emph {et~al.}}]{Seg14}%
  \BibitemOpen
  \bibfield  {author} {\bibinfo {author} {\bibfnamefont {J.}~\bibnamefont
  {Segovia}} \emph {et~al.},\ }\href {\doibase 10.1007/s00601-014-0907-2}
  {\bibfield  {journal} {\bibinfo  {journal} {Few-Body Systems}\ }\textbf
  {\bibinfo {volume} {55}},\ \bibinfo {pages} {1185} (\bibinfo {year}
  {2014})}\BibitemShut {NoStop}%
\bibitem [{\citenamefont {Segovia}\ \emph {et~al.}(2015)\citenamefont {Segovia}
  \emph {et~al.}}]{Seg15}%
  \BibitemOpen
  \bibfield  {author} {\bibinfo {author} {\bibfnamefont {J.}~\bibnamefont
  {Segovia}} \emph {et~al.},\ }\href {\doibase 10.1103/PhysRevLett.115.171801}
  {\bibfield  {journal} {\bibinfo  {journal} {Phys. Rev. Lett.}\ }\textbf
  {\bibinfo {volume} {115}},\ \bibinfo {pages} {171801} (\bibinfo {year}
  {2015})}\BibitemShut {NoStop}%
\bibitem [{\citenamefont {Clo\"{e}t}\ and\ \citenamefont
  {Roberts}(2014)}]{Cr14}%
  \BibitemOpen
  \bibfield  {author} {\bibinfo {author} {\bibfnamefont {I.~C.}\ \bibnamefont
  {Clo\"{e}t}}\ and\ \bibinfo {author} {\bibfnamefont {C.~D.}\ \bibnamefont
  {Roberts}},\ }\href {\doibase http://dx.doi.org/10.1016/j.ppnp.2014.02.001}
  {\bibfield  {journal} {\bibinfo  {journal} {Prog. Part. Nucl. Phys.}\
  }\textbf {\bibinfo {volume} {77}},\ \bibinfo {pages} {1 } (\bibinfo {year}
  {2014})}\BibitemShut {NoStop}%
\bibitem [{\citenamefont {Roberts}(2016)}]{Cr16}%
  \BibitemOpen
  \bibfield  {author} {\bibinfo {author} {\bibfnamefont {C.~D.}\ \bibnamefont
  {Roberts}},\ }\href {http://stacks.iop.org/1742-6596/706/i=2/a=022003}
  {\bibfield  {journal} {\bibinfo  {journal} {J. Phys. Conf. Ser.}\ }\textbf
  {\bibinfo {volume} {706}},\ \bibinfo {pages} {022003} (\bibinfo {year}
  {2016})}\BibitemShut {NoStop}%
\bibitem [{\citenamefont {{Roberts}}(2016)}]{Cr16a}%
  \BibitemOpen
  \bibfield  {author} {\bibinfo {author} {\bibfnamefont {C.~D.}\ \bibnamefont
  {{Roberts}}},\ }in\ \href {\doibase 10.7566/JPSCP.10.010012} {\emph {\bibinfo
  {booktitle} {Proceedings of the 10th International Workshop on the Physics of
  Excited Nucleons (NSTAR2015)}}}\ (\bibinfo {year} {2016})\ p.\ \bibinfo
  {pages} {010012}\BibitemShut {NoStop}%
\bibitem [{\citenamefont {Park}\ \emph {et~al.}(2015)\citenamefont {Park} \emph
  {et~al.}}]{Park15}%
  \BibitemOpen
  \bibfield  {author} {\bibinfo {author} {\bibfnamefont {K.}~\bibnamefont
  {Park}} \emph {et~al.} (\bibinfo {collaboration} {CLAS Collaboration}),\
  }\href {\doibase 10.1103/PhysRevC.91.045203} {\bibfield  {journal} {\bibinfo
  {journal} {Phys. Rev. C}\ }\textbf {\bibinfo {volume} {91}},\ \bibinfo
  {pages} {045203} (\bibinfo {year} {2015})}\BibitemShut {NoStop}%
\bibitem [{\citenamefont {Mokeev}\ \emph {et~al.}(2012)\citenamefont {Mokeev}
  \emph {et~al.}}]{Mo12}%
  \BibitemOpen
  \bibfield  {author} {\bibinfo {author} {\bibfnamefont {V.~I.}\ \bibnamefont
  {Mokeev}} \emph {et~al.} (\bibinfo {collaboration} {CLAS Collaboration}),\
  }\href {\doibase 10.1103/PhysRevC.86.035203} {\bibfield  {journal} {\bibinfo
  {journal} {Phys. Rev. C}\ }\textbf {\bibinfo {volume} {86}},\ \bibinfo
  {pages} {035203} (\bibinfo {year} {2012})}\BibitemShut {NoStop}%
\bibitem [{\citenamefont {Mokeev}\ and\ \citenamefont
  {Aznauryan}(2014)}]{Mo14}%
  \BibitemOpen
  \bibfield  {author} {\bibinfo {author} {\bibfnamefont {V.~I.}\ \bibnamefont
  {Mokeev}}\ and\ \bibinfo {author} {\bibfnamefont {I.~G.}\ \bibnamefont
  {Aznauryan}},\ }\href {\doibase 10.1142/S2010194514600805} {\bibfield
  {journal} {\bibinfo  {journal} {J. of Phys. Conf. Ser.}\ }\textbf {\bibinfo
  {volume} {26}},\ \bibinfo {pages} {1460080} (\bibinfo {year}
  {2014})}\BibitemShut {NoStop}%
\bibitem [{\citenamefont {Mokeev}\ \emph
  {et~al.}(2016{\natexlab{b}})\citenamefont {Mokeev} \emph {et~al.}}]{Mo16a}%
  \BibitemOpen
  \bibfield  {author} {\bibinfo {author} {\bibfnamefont {V.~I.}\ \bibnamefont
  {Mokeev}} \emph {et~al.},\ }\href {\doibase 10.1051/epjconf/201611301013}
  {\bibfield  {journal} {\bibinfo  {journal} {EPJ Web Conf.}\ }\textbf
  {\bibinfo {volume} {113}},\ \bibinfo {pages} {01013} (\bibinfo {year}
  {2016}{\natexlab{b}})}\BibitemShut {NoStop}%
\bibitem [{\citenamefont {Mokeev}(2016)}]{Mo16b}%
  \BibitemOpen
  \bibfield  {author} {\bibinfo {author} {\bibfnamefont {V.~I.}\ \bibnamefont
  {Mokeev}},\ }\href {\doibase 10.1007/s00601-016-1127-8} {\bibfield  {journal}
  {\bibinfo  {journal} {Few-Body Systems}\ }\textbf {\bibinfo {volume} {57}},\
  \bibinfo {pages} {909} (\bibinfo {year} {2016})}\BibitemShut {NoStop}%
\bibitem [{\citenamefont {Ripani}\ \emph {et~al.}(2003)\citenamefont {Ripani}
  \emph {et~al.}}]{Ri03}%
  \BibitemOpen
  \bibfield  {author} {\bibinfo {author} {\bibfnamefont {M.}~\bibnamefont
  {Ripani}} \emph {et~al.} (\bibinfo {collaboration} {CLAS Collaboration}),\
  }\href {\doibase 10.1103/PhysRevLett.91.022002} {\bibfield  {journal}
  {\bibinfo  {journal} {Phys. Rev. Lett.}\ }\textbf {\bibinfo {volume} {91}},\
  \bibinfo {pages} {022002} (\bibinfo {year} {2003})}\BibitemShut {NoStop}%
\bibitem [{\citenamefont {Anisovich}\ \emph {et~al.}(2012)\citenamefont
  {Anisovich} \emph {et~al.}}]{Sar12}%
  \BibitemOpen
  \bibfield  {author} {\bibinfo {author} {\bibfnamefont {A.~V.}\ \bibnamefont
  {Anisovich}} \emph {et~al.},\ }\href {\doibase 10.1140/epja/i2012-12015-8}
  {\bibfield  {journal} {\bibinfo  {journal} {Eur. Phys. J. A}\ }\textbf
  {\bibinfo {volume} {48}},\ \bibinfo {pages} {15} (\bibinfo {year}
  {2012})}\BibitemShut {NoStop}%
\bibitem [{\citenamefont {Fedotov}\ \emph {et~al.}(2009)\citenamefont {Fedotov}
  \emph {et~al.}}]{Fe09}%
  \BibitemOpen
  \bibfield  {author} {\bibinfo {author} {\bibfnamefont {G.~V.}\ \bibnamefont
  {Fedotov}} \emph {et~al.} (\bibinfo {collaboration} {CLAS Collaboration}),\
  }\href {\doibase 10.1103/PhysRevC.79.015204} {\bibfield  {journal} {\bibinfo
  {journal} {Phys. Rev. C}\ }\textbf {\bibinfo {volume} {79}},\ \bibinfo
  {pages} {015204} (\bibinfo {year} {2009})}\BibitemShut {NoStop}%
\bibitem [{\citenamefont {Mokeev}\ \emph {et~al.}(2009)\citenamefont {Mokeev}
  \emph {et~al.}}]{Mo09}%
  \BibitemOpen
  \bibfield  {author} {\bibinfo {author} {\bibfnamefont {V.~I.}\ \bibnamefont
  {Mokeev}} \emph {et~al.},\ }\href {\doibase 10.1103/PhysRevC.80.045212}
  {\bibfield  {journal} {\bibinfo  {journal} {Phys. Rev. C}\ }\textbf {\bibinfo
  {volume} {80}},\ \bibinfo {pages} {045212} (\bibinfo {year}
  {2009})}\BibitemShut {NoStop}%
\bibitem [{\citenamefont {Anisovich}\ \emph {et~al.}(2014)\citenamefont
  {Anisovich} \emph {et~al.}}]{Sar14}%
  \BibitemOpen
  \bibfield  {author} {\bibinfo {author} {\bibfnamefont {A.~V.}\ \bibnamefont
  {Anisovich}} \emph {et~al.},\ }\href {\doibase 10.1140/epja/i2014-14129-3}
  {\bibfield  {journal} {\bibinfo  {journal} {Eur. Phys. J. A}\ }\textbf
  {\bibinfo {volume} {50}},\ \bibinfo {pages} {129} (\bibinfo {year}
  {2014})}\BibitemShut {NoStop}%
\bibitem [{\citenamefont {Kamano}\ \emph {et~al.}(2013)\citenamefont {Kamano}
  \emph {et~al.}}]{Lee13}%
  \BibitemOpen
  \bibfield  {author} {\bibinfo {author} {\bibfnamefont {H.}~\bibnamefont
  {Kamano}} \emph {et~al.},\ }\href {\doibase 10.1103/PhysRevC.88.035209}
  {\bibfield  {journal} {\bibinfo  {journal} {Phys. Rev. C}\ }\textbf {\bibinfo
  {volume} {88}},\ \bibinfo {pages} {035209} (\bibinfo {year}
  {2013})}\BibitemShut {NoStop}%
\bibitem [{\citenamefont {Kamano}\ \emph {et~al.}(2016)\citenamefont {Kamano}
  \emph {et~al.}}]{Lee15}%
  \BibitemOpen
  \bibfield  {author} {\bibinfo {author} {\bibfnamefont {H.}~\bibnamefont
  {Kamano}} \emph {et~al.},\ }\href {\doibase 10.1103/PhysRevC.94.015201}
  {\bibfield  {journal} {\bibinfo  {journal} {Phys. Rev. C}\ }\textbf {\bibinfo
  {volume} {94}},\ \bibinfo {pages} {015201} (\bibinfo {year}
  {2016})}\BibitemShut {NoStop}%
\bibitem [{\citenamefont {Mecking}\ \emph {et~al.}(2003)\citenamefont {Mecking}
  \emph {et~al.}}]{Me03}%
  \BibitemOpen
  \bibfield  {author} {\bibinfo {author} {\bibfnamefont {B.~A.}\ \bibnamefont
  {Mecking}} \emph {et~al.},\ }\href {\doibase
  http://dx.doi.org/10.1016/S0168-9002(03)01001-5} {\bibfield  {journal}
  {\bibinfo  {journal} {Nucl. Instr. and Meth.}\ }\textbf {\bibinfo {volume}
  {503}},\ \bibinfo {pages} {513 } (\bibinfo {year} {2003})}\BibitemShut
  {NoStop}%
\bibitem [{\citenamefont {Golovach}()}]{Gol12}%
  \BibitemOpen
  \bibfield  {author} {\bibinfo {author} {\bibfnamefont {E.~N.}\ \bibnamefont
  {Golovach}},\ }\href@noop {} {}\bibinfo {howpublished}
  {\url{http://depni.sinp.msu.ru/~golovach/EG/}}\BibitemShut {NoStop}%
\bibitem [{\citenamefont {Mo}\ and\ \citenamefont {Tsai}(1969)}]{Mo69}%
  \BibitemOpen
  \bibfield  {author} {\bibinfo {author} {\bibfnamefont {L.~W.}\ \bibnamefont
  {Mo}}\ and\ \bibinfo {author} {\bibfnamefont {Y.~S.}\ \bibnamefont {Tsai}},\
  }\href {\doibase 10.1103/RevModPhys.41.205} {\bibfield  {journal} {\bibinfo
  {journal} {Rev. Mod. Phys.}\ }\textbf {\bibinfo {volume} {41}},\ \bibinfo
  {pages} {205} (\bibinfo {year} {1969})}\BibitemShut {NoStop}%
\bibitem [{\citenamefont {Aznauryan}\ \emph {et~al.}(2005)\citenamefont
  {Aznauryan} \emph {et~al.}}]{Mo06}%
  \BibitemOpen
  \bibfield  {author} {\bibinfo {author} {\bibfnamefont {I.~G.}\ \bibnamefont
  {Aznauryan}} \emph {et~al.},\ }\href {\doibase 10.1103/PhysRevC.72.045201}
  {\bibfield  {journal} {\bibinfo  {journal} {Phys. Rev. C}\ }\textbf {\bibinfo
  {volume} {72}},\ \bibinfo {pages} {045201} (\bibinfo {year}
  {2005})}\BibitemShut {NoStop}%
\bibitem [{\citenamefont {Mokeev}\ \emph {et~al.}(2001)\citenamefont {Mokeev}
  \emph {et~al.}}]{Mo01}%
  \BibitemOpen
  \bibfield  {author} {\bibinfo {author} {\bibfnamefont {V.~I.}\ \bibnamefont
  {Mokeev}} \emph {et~al.},\ }\href {\doibase 10.1134/1.1389557} {\bibfield
  {journal} {\bibinfo  {journal} {Phys. Atom. Nucl.}\ }\textbf {\bibinfo
  {volume} {64}},\ \bibinfo {pages} {1292} (\bibinfo {year}
  {2001})}\BibitemShut {NoStop}%
\bibitem [{\citenamefont {Ripani}\ \emph {et~al.}(2000)\citenamefont {Ripani}
  \emph {et~al.}}]{Ri00}%
  \BibitemOpen
  \bibfield  {author} {\bibinfo {author} {\bibfnamefont {M.}~\bibnamefont
  {Ripani}} \emph {et~al.},\ }\href {\doibase
  http://dx.doi.org/10.1016/S0375-9474(99)00853-2} {\bibfield  {journal}
  {\bibinfo  {journal} {Nucl. Phys. A}\ }\textbf {\bibinfo {volume} {672}},\
  \bibinfo {pages} {220 } (\bibinfo {year} {2000})}\BibitemShut {NoStop}%
\bibitem [{\citenamefont {Bosted}(1995)}]{Bost}%
  \BibitemOpen
  \bibfield  {author} {\bibinfo {author} {\bibfnamefont {P.~E.}\ \bibnamefont
  {Bosted}},\ }\href {\doibase 10.1103/PhysRevC.51.409} {\bibfield  {journal}
  {\bibinfo  {journal} {Phys. Rev. C}\ }\textbf {\bibinfo {volume} {51}},\
  \bibinfo {pages} {409} (\bibinfo {year} {1995})}\BibitemShut {NoStop}%
\bibitem [{\citenamefont {Fedotov}()}]{resnum}%
  \BibitemOpen
  \bibfield  {author} {\bibinfo {author} {\bibfnamefont {G.~V.}\ \bibnamefont
  {Fedotov}},\ }\href@noop {} {}\bibinfo {howpublished}
  {\url{https://userweb.jlab.org/~mokeev/resonance_electrocouplings/}}\BibitemShut
  {NoStop}%
\bibitem [{\citenamefont {Isupov}(2017)}]{Is16}%
  \BibitemOpen
  \bibfield  {author} {\bibinfo {author} {\bibfnamefont {E.~L.}\ \bibnamefont
  {Isupov}},\ }\href@noop {} {}\bibinfo {howpublished} {See Supplemental
  Material at [URL will be inserted by publisher]} (\bibinfo {year}
  {2017})\BibitemShut {NoStop}%
\bibitem [{\citenamefont {Denizli}\ \emph {et~al.}(2007)\citenamefont {Denizli}
  \emph {et~al.}}]{Den07}%
  \BibitemOpen
  \bibfield  {author} {\bibinfo {author} {\bibfnamefont {H.}~\bibnamefont
  {Denizli}} \emph {et~al.} (\bibinfo {collaboration} {CLAS Collaboration}),\
  }\href {\doibase 10.1103/PhysRevC.76.015204} {\bibfield  {journal} {\bibinfo
  {journal} {Phys. Rev. C}\ }\textbf {\bibinfo {volume} {76}},\ \bibinfo
  {pages} {015204} (\bibinfo {year} {2007})}\BibitemShut {NoStop}%
\bibitem [{\citenamefont {{Particle Data Group, K.~A. Olive}}\ \emph
  {et~al.}(2014)\citenamefont {{Particle Data Group, K.~A. Olive}} \emph
  {et~al.}}]{Rpp14}%
  \BibitemOpen
  \bibfield  {author} {\bibinfo {author} {\bibnamefont {{Particle Data Group,
  K.~A. Olive}}} \emph {et~al.},\ }\href
  {http://stacks.iop.org/1674-1137/38/i=9/a=090001} {\bibfield  {journal}
  {\bibinfo  {journal} {Chinese Physics C}\ }\textbf {\bibinfo {volume} {38}},\
  \bibinfo {pages} {090001} (\bibinfo {year} {2014})}\BibitemShut {NoStop}%
\bibitem [{\citenamefont {Santopinto}\ and\ \citenamefont
  {Giannini}(2015)}]{San15}%
  \BibitemOpen
  \bibfield  {author} {\bibinfo {author} {\bibfnamefont {E.}~\bibnamefont
  {Santopinto}}\ and\ \bibinfo {author} {\bibfnamefont {M.~M.}\ \bibnamefont
  {Giannini}},\ }\href@noop {} {\bibfield  {journal} {\bibinfo  {journal}
  {Chin. J. Phys.}\ }\textbf {\bibinfo {volume} {53}},\ \bibinfo {pages}
  {020301} (\bibinfo {year} {2015})}\BibitemShut {NoStop}%
\end{thebibliography}%
\bibliographystyle{apsrev4-1}
\end{document}